\documentstyle[12pt,a4,bbbold,psfig]{article}

\newcommand{\grad}{\mathop{\rm grad}\nolimits}
\def\dfrac#1#2{{\displaystyle#1\over\displaystyle#2}}
\def\slantfrac#1#2{\hbox{$\,^#1\!/_#2$}}

\begin{document}

\title{Three-dimensional modeling of mass transfer in close
binary systems with non-synchronous rotation}
\author{
Bisikalo D.V.$^1$, Boyarchuk A.A.$^1$,\\
Kuznetsov O.A.$^2$, Chechetkin V.M.$^2$\\[0.3cm]
$^1$ {\it Institute of Astronomy of the Russian Acad. of Sci.,
Moscow}\\
{\sf bisikalo@inasan.rssi.ru; aboyar@inasan.rssi.ru}\\[0.3cm]
$^2$ {\it Keldysh Institute of Applied Mathematics, Moscow}\\
{\sf kuznecov@spp.keldysh.ru; chech@int.keldysh.ru}\\[0.3cm]
}
\date{}
\maketitle

\begin{abstract}
{\bf Abstract}---We present the results of three-dimensional
numerical simulations of mass transfer in semi-detached binary
systems in which the mass-losing star is rotating.  The cases
of aligned and misaligned non-synchronous rotation of the donor
star are considered; the resulting flow patterns are compared to
the synchronous case. The main properties of the
flow, such as the formation of an circumbinary envelope, the
absence of a "hot spot" on the edge of the accretion disk, and
the formation of a shock wave along the flow edge, are
qualitatively similar to those obtained earlier. For the case of
misaligned, non-synchronous rotation, the behavior of the disk and
surrounding matter in established flow regime reflects
changes in the boundary conditions at the surface of the donor
star; in other words, a "driven disk" model is realized in the
calculations.

\end{abstract}

\section*{INTRODUCTION}
   Since the discovery of non-synchronous rotation (when the
rotational velocity of one or both components differs from the
orbital angular velocity of the system) by Schlesinger in the
spectral binary $\delta$~Lib in 1909 and in the eclipsing
binary $\lambda$~Tau in 1910 [1], non-synchronous rotation has
been detected in numerous binary systems of various types. For a
number of Algol-type systems (RZ~Sct, U~Cep, RY~Per), there is
evidence for the rapid rotation of one of the components [2, 3].
According to [2, 3], the degree to which the rotation is
non-synchronous $f=\Omega_\star/\Omega$ is rather high for these
systems: $f=6.6 \div 9.0$ for RZ~Sct and $f=10.8 \div 14.5$ for
RY~Per. The presence of rotation for one component that differs
from the orbital rotational velocity has also been detected for
several long-period RS CVn stars (TZ~For,
$\lambda$~And, AY~Cet [4] and $\alpha$~Aur (Capella) [5, 6]).
The estimated rotation non-synchronicity of the hot star in the
$\alpha$~Aur system reaches $f\sim 10\div 12$ [7--9].

   For a number of binary systems, various observational
properties (light curves, radial velocity curves, etc) also
appear to show long-period variations on timescales
substantially longer than the orbital period.  Among the best
known binaries of this type are the systems Her X-1 (HZ Her) and
SS433. In order to explain the long-period variations in the
Her X-1 system,  a "driven disk" model was suggested in
[10--12], based on the idea that the rotation of the donor star
is non-synchronous and misaligned.  A similar model was also
considered later for SS433 [13--17].

   The first attempts to study the role of non-synchronicity in
mass transfer were made in a ballistic approximation [9, 18].
However, the fact that gas-dynamical factors were not taken into
account in these calculations casts doubt on the reliability of
the results. Due to the three-dimensional nature of the problem,
a correct analysis of the flow pattern in systems with
non-synchronous rotation is possible only in the framework of 3D
gas-dynamical models. Model calculations for binary systems with
non-synchronously rotating components were presented in [19, 20].
Unfortunately, these studies used a simplified formulation of
the problem, and did not take into account the change of the
shape of the donor star in the case of non-synchronous rotation
[21]; therefore, the results obtained require additional
verification. In addition, the solution given in [19, 20] was
limited, since the impact of the circumbinary gas on the flow
structure near the donor star was not taken into consideration.
Calculations made later [22--25] for the case of synchronous
rotation indicated that the influence of the circumbinary gas
can substantially change the structure of gaseous flows in the
vicinity of $L_1$ in a steady-state flow regime.  Here, we
present results for our solutions to this problem formulated in
a self-consistent way, devoid of the drawbacks of previous
models.

\section{THE MODEL}

\subsection{Binary system parameters}

   In our previous studies [22, 23], we considered the flow
morphology in a low-mass X-ray binary system with synchronously
rotating components in a three-dimensional formulation. In order
to study the impact of non-synchronous rotation of the donor star
on the flow structure, we now consider a semi-detached system
with the same parameters as those in [22, 23]. We adopted
typical parameters for a low-mass X-ray binary, close to these
of X1822--371 [26].  The primary component was assumed to fill
its Roche lobe and to have a mass $M_1=0.28M_\odot$
and surface temperature $T=10^4$~K; the mass $M_2$ of
the secondary, a compact object with radius $0.05R_\odot$, was
assumed to be $1.4M\odot$. The orbital period of the system was
$P_{orb} = 5^h.56$, and the distance between the centers of
the components was $A = 1.97 R_\odot$.

   We will use the following terminology. We will call
"synchronous" the case when all periods---i.e., the rotation
periods of both of the stars and of the system as a
whole---coincide ($P_{\star1}=P_{\star2}=P_{orb}$);
"non-synchronous" rotation is the case when the rotation period of
the donor star does not coincide with the orbital rotation
period. For non-synchronous rotation of the donor star, two
positions of its rotation axis relative to the
system are possible: 1) aligned, when the rotation axis of the
donor star is perpendicular to the orbital plane of the binary
system; and 2) misaligned, when this axis is inclined to this
direction. For all calculations, we will assume that the
rotation axis of the donor star goes through its center of mass
and is fixed in the laboratory frame.

   We assumed that the rotation velocity of the donor star in
the laboratory frame is twice the orbital rotation velocity of
the system. Three-dimensional calculations were performed for
both aligned and misaligned rotation of the donor star.

\subsection{Shape of the mass-losing star}

   The shape of the mass-losing component in a
semi-detached binary system can easily be determined for the
standard case, when it is assumed that: (i) the orbits of the
components are circular, (ii) the rotation of the star is
synchronized with the orbital rotation ${\bmath \Omega}_\star={\bmath
\Omega}$, and (iii) the stars are strongly concentrated, so that
their gravitational fields can be considered to be the fields of
point masses. Under these conditions, in the adopted coordinate
frame (the $X$ axis is directed along the line connected the
centers of the stars, the $Z$ axis coincides with the rotation
axis, and the $Y$ axis forms a right-handed coordinate system
whose origin is at the center of mass of the donor star), the
total potential in the Roche approximation can be written

$$
\Phi({\bmath r}) = -\frac{G M_1}{d_1}-\frac{G M_2}{d_2}
-\frac{1}{2}\Omega^2\left((x-x_c)^2+y^2\right)\eqno(1)
$$
and the shape of the donor star coincides with its Roche lobe;
i.e., with the equipotential surface passing through the inner
Lagrange point $L_1$ (see, for example, [27, 28]). When the star
reaches the boundary of its Roche lobe, mass transfer begins
through $L_1$, where the pressure gradient is not
balanced by other forces. In (1), $M_1$ and
$M_2$ are the masses of the components, $M=M_1+M_2$ is the total
mass of the system, $\Omega$ is the orbital angular velocity of
the system, ${\bmath\Omega}=(0,0,\Omega)$; $x_c=A M_2/M$;
$d_1=\sqrt{x^2+y^2+z^2}$ is the distance to the center of mass
of the primary (the donor star), $d_2=\sqrt{(x-A)^2+y^2+z^2}$ is
the distance to the center of mass of the accretor, and $A$ is
the distance between the components.

   The situation changes when the Roche approximation is not
satisfied, so that the potential has a shape that differs from
(1). Let us consider the general case, when the mass-losing
component rotates non-synchronously and its rotation is misaligned
with the orbital rotation:
${\bmath\Omega}_\star~\lefteqn{\parallel}{/}~{\bmath \Omega}$.
Following [29, 30], where the potential for the case of
non-synchronous rotation is considered, and [21, 31, 32], where
misaligned rotation is taken into account, we will determine
the position of the Lagrange points, where the pressure gradient
is not compensated by other forces.  Assuming that ${\bmath
\Omega}_\star$ does not change with time (there is no
precession) and that the rotation of the star is uniform [21,
33], we can write the equation of motion for a test particle in
a reference frame rotating with angular velocity
${\bmath\Omega}_\star$ in the form (without account for the
pressure force):

$$
\ddot{\bmath r}=
-\grad\left(-\frac{G M_1}{d_1}-\frac{GM_2}{d_2}\right)
-\ddot{\bmath r}_0
-{\bmath\Omega}_\star\times\left[{\bmath \Omega}_\star\times{\bmath
r}\right]
-2\,{\bmath \Omega}_\star\times\dot{\bmath r}\,,
$$
where $\ddot{\bmath r}_0$ denotes the acceleration of the coordinate origin (the donor
star's center of gravity) in the laboratory frame:

$$
\ddot{\bmath r}_0={\bmath\Omega}\times\left[{\bmath \Omega}\times
(-{\bmath r}_c)\right]=(\Omega^2x_c,0,0)
$$

$$
{\bmath r}_c=(x_c,0,0)\,.
$$
We obtain after simple manipulation

$$
\ddot{\bmath r}={\bmath F}
$$

$$
{\bmath F}=-\grad\left(-\frac{G M_1}{d_1}-\frac{GM_2}{d_2}\right)
-{\bmath\Omega}\times\left[{\bmath \Omega}\times
(-{\bmath r}_c)\right]
-{\bmath\Omega}_\star\times\left[{\bmath \Omega}_\star\times{\bmath
r}\right]
-2\,{\bmath \Omega}_\star\times\dot{\bmath r}\,.
$$

\renewcommand{\thefigure}{1}
\begin{figure}[t]
\centerline{\hbox{\psfig{figure=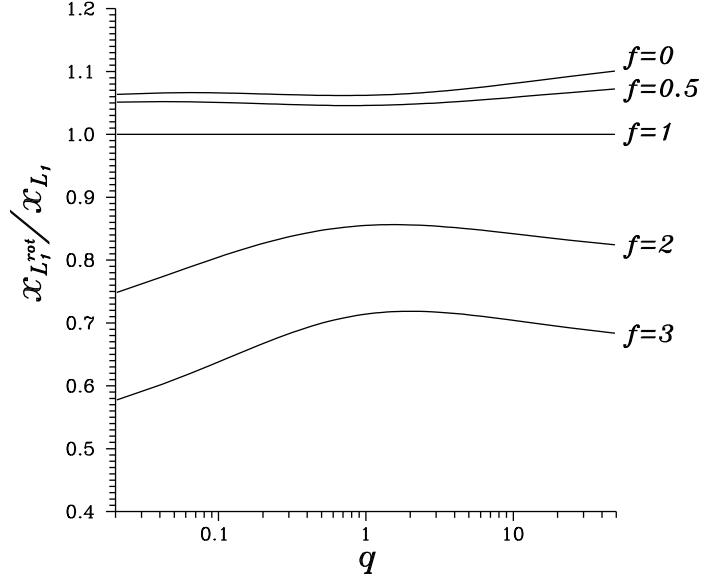,height=3.0in}}}
\caption{\sl
Relative change of the position of the inner Lagrange point for
systems with non-synchronous rotation $x_{L_1^{rot}}/x_{L_1}$as a
function of the mass ratio of the components $q=M_2/M_1$ for
various values of the degree of non-synchronicity of the rotation
$f=\Omega_\star/\Omega$.
}
\end{figure}

   The positions of the Lagrange points can be determined from
the condition ${\bmath F}=0$. When there exists a potential $\Psi$
such that ${\bmath F}=-\grad\Psi$, these points are stationary
points of the potential.  Using the relations

$$
-{\bmath\Omega}\times\left[{\bmath \Omega}\times
(-{\bmath r}_c)\right]=(-\Omega^2x_c,0,0)
=-\grad\left({\bmath \Omega}^2 x_c x\right)
$$

$$
-{\bmath\Omega}_\star\times\left[{\bmath \Omega}_\star\times{\bmath
r}\right]=-\grad\left(-\frac{1}{2}
{\bmath \Omega}_\star^2\Delta^2\right)\,,
$$
where $\Delta$ is the distance from the point ${\bmath r}$ to the
rotation axis ${\bmath \Omega}_\star$, we find that the desired
potential $\Psi$ has the form

$$
\Psi({\bmath r})=
-\frac{G M_1}{d_1}-\frac{GM_2}{d_2}
+{\bmath \Omega}^2 x_c x
-\frac{1}{2}
{\bmath \Omega}_\star^2\Delta^2\,,
\eqno(2)
$$
and the equation of motion can now be represented

$$
\ddot{\bmath r}
=-\grad\Psi
-2\,{\bmath \Omega}_\star\times\dot{\bmath r}\,.
$$
The potential $\Psi$ includes all forces acting in the rotating
reference frame, except for the Coriolis force and the pressure
gradient.  For the special case of aligned rotation
(${\bmath \Omega}_{\star x}={\bmath \Omega}_{\star y}=0$)
$\Delta^2=x^2+y^2$
and the potential (to within a constant) is
described by the expression

$$
\Psi({\bmath r})=
-\frac{G M_1}{d_1}-\frac{GM_2}{d_2}
-\frac{1}{2}{\bmath \Omega}^2 \left((x-x_c)^2+y^2\right)
-\frac{1}{2}
({\bmath \Omega}_\star^2-{\bmath \Omega}^2)\left(x^2+y^2\right)\,.
\eqno(3)
$$
Note that, in the case of aligned, synchronous rotation
(${\bmath \Omega}_{\star}={\bmath \Omega}$), the
potential coincides with (1).

   Using relations (1) and (3), we can determine the change of
the position of the inner Lagrange point $L_1^{rot}$ for the
case of aligned, non-synchronous rotation compared to the
standard case.  For the sake of convenience, we will write the
standard Roche potential (1) and the potential for aligned,
non-synchronous rotation (3) in dimensionless form (the distances
are divided by $A$, the potential is divided by $GM_1/A$, and $q
= M_2/M_1$):

$$
\Phi({\bmath r})=-\frac{1}{\sqrt{x^2+y^2+z^2}}
-\frac{q}{\sqrt{(x-1)^2+y^2+z^2}}
-\frac{1}{2}(q+1)\left((x-x_c)^2+y^2\right)\,,
$$

$$
\Psi({\bmath r})=-\frac{1}{\sqrt{x^2+y^2+z^2}}
-\frac{q}{\sqrt{(x-1)^2+y^2+z^2}}
-\frac{1}{2}(q+1)\left((x-x_c)^2+y^2\right)
$$
$$
~~~~~~~~~~~~~~~~~~-\frac{1}{2}(q+1)(f^2-1)\left(x^2+y^2\right)\,.
$$
Figure 1 presents the dependencies of $x_{L_1^{rot}}/x_{L_1}$
on $q=M_2/M_1$ and $f=\Omega_\star/\Omega$. We can see that,
when the rotation of the star is slower than the orbital
rotation ($f<1$), the Roche lobe constructed with account for
non-synchronous rotation is larger than the standard Roche lobe,
with its maximum size achieved when $f=0$. When the rotation of
the star is faster than the orbital rotation, the "non-synchronous"
Roche lobe is smaller than the standard Roche lobe, with
$x_{L_1^{rot}}/x_{L_1} \to 0$ for $f\to\infty$.

   For the case of misaligned rotation, the expression for
potential (2) becomes very complicated. We will specify the
position of the vector ${\bmath\Omega}_\star$ by two angles: the
angle $\vartheta$ between ${\bmath\Omega}_\star$ and the $Z$ axis,
and the angle $\phi$ between the $X$ axis and the projection of
${\bmath \Omega}_\star$ onto the $XY$ plane. The dimensionless
potential can then be written

$$
\Psi(x,y,z)=-\frac{1}{\sqrt{x^2+y^2+z^2}}
-\frac{q}{\sqrt{(x-1)^2+y^2+z^2}}
$$
$$
~~~~~~~~~~~+q(q+1)x-\frac{1}{2}(q+1)f^2\Delta^2(x,y,z)
\eqno(4)
$$

$$
\Delta^2(x,y,z)=
x^2(1-\cos^2\phi\sin^2\vartheta)+y^2(1-\sin^2\phi\sin^2\vartheta)
+z^2\sin^2\vartheta
$$
$$
-xy\sin^2\vartheta\sin2\phi-xz\cos\phi\sin2\vartheta
-yz\sin\phi\sin2\vartheta\,.
$$
In the case considered, the potential for the misaligned
rotation depends on four parameters: $q$, $f$, $\vartheta$, and
$\phi$.  The last parameter changes with time, since the vector
${\bmath \Omega}_\star$ rotates with angular velocity $-\Omega$ in
a reference frame rotated with the binary system.  The inner
Lagrange point no longer lies on the line connecting the centers
of the components.  Moreover, according to [21], for some
parameter values, the potential at the outer Lagrange point
$L_2$ can turn out to be lower than that at the inner Lagrange
point.  In this case, the Roche lobe will "open" from the side
of $L_2$ earlier, than it does from the side of $L_1$.  However,
as this occurs for large values of $f$ and $\vartheta$ (see Fig.
5 from [21]), we will not consider this possibility in our
study.

   Let us estimate how much the position of $L_1$ deviates from
the line connecting the component centers for the case of
misaligned rotation.  Given that, first,
$|y_{L_1^{rot}}|<|z_{L_1^{rot}}|$ for not very large
$\vartheta$, and, second, the maximum of $|z_{L_1^{rot}}|$ is
reached when $\phi=0$, we will estimate the deviation of
$L_1^{rot}$ from the line connecting the component centers using
the value $|z_{L_1^{rot}}|$ for $\phi=0$. Our analysis of the
vertical shift of the inner Lagrange point indicates that, for
the adopted parameters, this deviation is negligible, since it
is smaller than the cross section of the flow determined using
standard models [34, 28].

   This fact made it possible to simplify the model, and to
assume that the shape of the star does not change as a function
of orbital phase.  The shape of the donor star is taken to be
that of the equipotential surface (4) passing through
$L_1^{rot}$. Since, for $\phi=\pi/2$, the inner Lagrange point
$L_1^{rot}$ lies along the line connecting the component
centers, we used expression (4) for the potential with the fixed
value $\phi=\pi/2$ to determine the shape of the donor star.

\subsection{Gasdynamical equations}

   The flow of gas in a binary system is described by a system of
gas-dynamical equations:

$$
\begin{array}{c}
\dfrac{\partial \rho}{\partial t}
+\dfrac{\partial \rho u}{\partial x}
+\dfrac{\partial \rho v}{\partial y}
+\dfrac{\partial \rho w}{\partial z}=0\\
~\\
\dfrac{\partial \rho u}{\partial t}
+\dfrac{\partial (\rho u^2+P)}{\partial x}
+\dfrac{\partial \rho uv}{\partial y}
+\dfrac{\partial \rho uw}{\partial z}
= -\rho\dfrac{\partial\Phi}{\partial x}+2\Omega v\rho\\
~\\
\dfrac{\partial \rho v}{\partial t}
+\dfrac{\partial \rho uv}{\partial x}
+\dfrac{\partial (\rho v^2+P)}{\partial y}
+\dfrac{\partial \rho vw}{\partial z}
= -\rho\dfrac{\partial\Phi}{\partial y}-2\Omega u\rho\\
~\\
\dfrac{\partial \rho w}{\partial t}
+\dfrac{\partial \rho uw}{\partial x}
+\dfrac{\partial \rho vw}{\partial y}
+\dfrac{\partial (\rho w^2+P)}{\partial z}
= -\rho\dfrac{\partial\Phi}{\partial z}\\
~\\
\dfrac{\partial \rho E}{\partial t}
+\dfrac{\partial \rho uh}{\partial x}
+\dfrac{\partial \rho vh}{\partial y}
+\dfrac{\partial \rho wh}{\partial z}
= -\rho u\dfrac{\partial\Phi}{\partial x}
-\rho v\dfrac{\partial\Phi}{\partial y}
-\rho w\dfrac{\partial\Phi}{\partial z}\,.\\
~
\end{array}\eqno(5)
$$
Here, $\rho$ denotes density; $u$, $v$, and $w$ are the $x$,
$y$, and $z$ components of the velocity vector ${\bmath
v}=(u,v,w)$; $P$ is the pressure;
$E=\varepsilon+\slantfrac{1}{2}\cdot|{\bmath v}|^2$ is the
total specific energy;
$h=\varepsilon+P/\rho+\slantfrac{1}{2}\cdot|{\bmath v}|^2$ is the
total specific enthalpy; and $\Phi$ is the Roche potential. The
gas-dynamical equations are written in a reference frame that
rotates with the binary system, so that the Roche potential (1)
is used in (5).  To close the system of equations (5), we used
the equation of state for an ideal gas,
$P=(\gamma-1)\rho\varepsilon$.  To take into account radiative
losses, we took the adiabatic index $\gamma$ to be 1.01 [35,
36].  The non-synchronicity of the donor-star rotation was taken
into account when specifying the boundary conditions.

\subsection{Numerical model}

   To solve the above system of equations, we used the TVD
scheme of Roe with a high approximation order [22, 37, 38]. The
system of equations was solved from arbitrarily chosen initial
conditions to the point when the flow regime became
steady-state. The calculations for the case of aligned rotation
included five orbital periods; the solution was stationary over
the last three periods. In the case of misaligned rotation, it
is impossible to achieve a true steady-state regime, due to the
periodic time dependence of the boundary conditions. Therefore,
we took the solution to be steady-state when the main features
of the flow structure repeated with a period equal to that of
the boundary conditions. We made calculations over six orbital
periods, though strict periodicity of the solution was already
seen for times longer than two orbital periods. The calculation
area was the parallelepiped
$[-A...2A]\times[-A...A]\times[-A...A]$.  We used a nonuniform
grid consisting of $91\times 81\times 55$ nodes that were more
densely spaced in the accretor zone.

   We adopted free-outflow conditions for the matter at the
accretor and at the outer boundary of the calculation zone. We
used (4) to determine the shape of the donor star, which
coincided with the equipotential surface passing through the
inner Lagrange point $L_1^{rot}$ for $\phi=\pi/2$. The boundary
conditions at the surface of this star were determined by
solving for the decay of the discontinuity between the gas
parameters ($\rho_0,~{\bmath v}_0,~P_0$) at the surface of the
mass-losing star and in the calculation cell that was nearest
to the given point of the surface [35]. Note that the boundary
value of the density does not affect the solution, since the
system of equations scales in $\rho$ and $P$. In the
calculations, we chose an arbitrary value of $\rho_0$; to
determine the real densities in a specific system with a known
mass-loss rate, the calculated densities must simply be
increased in accordance with a scale determined from the ratio
of the real and model densities at the surface of the
mass-losing component.

   The velocity at the surface of the donor star was specified as

$$
{\bmath v}_0={\bmath \Omega}_\star^{rot} \times {\bmath r}+{\bmath n}
c_0\,,
$$
where
${\bmath\Omega}_\star^{rot}={\bmath\Omega}_\star-{\bmath\Omega}$
is the angular-velocity vector for the rotation of the donor
star in the rotating reference frame, $c_0$ is the sound speed at the
stellar surface, and ${\bmath n}$ is the normal vector to the stellar surface.
Note that our assumption that the star's rotation is uniform
implies that the velocity vector at the surface of the donor star has a
component normal to the surface, which is due, not only to the sound speed,
but also to the rotation of the star. It is evident that, in
the equilibrium state for the stellar surface, no non-thermal movement of
gas can occur normal to the surface.

\renewcommand{\thefigure}{2}
\begin{figure}[t]
\centerline{\hbox{\psfig{figure=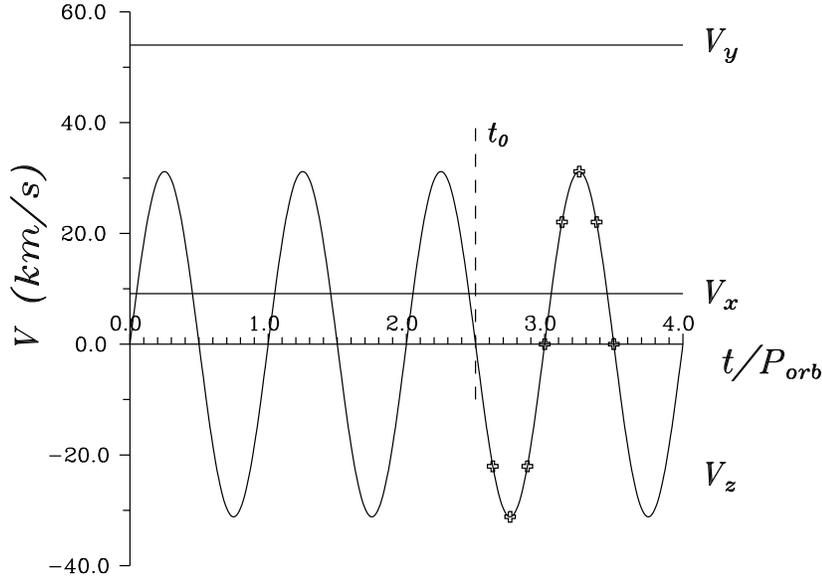,height=3.0in}}}
\caption{\sl
Components of velocity vector at the inner Lagrange point for
various cases: (1) synchronous rotation $V_x=c_0$, $V_y=0$,
$V_z=0$; (2) aligned, non-synchronous rotation $V_x=c_0$,
$V_y=x_{L_1^{rot}}\Omega^{rot}_{\star}$, $V_z=0$; (3)
misaligned, non-synchronous rotation for $\vartheta^*=30^\circ$
(the angle between the rotation vector in the rotating reference
frame ${\bmath \Omega}_\star^{rot}$ and the $Z$ axis):  $V_x=c_0$,
$V_y=x_{L_1^{rot}}\Omega^{rot}_{\star}\cos(\vartheta^*)$,
$V_z=-x_{L_1^{rot}}\Omega^{rot}_{\star}\sin(\vartheta^*)\sin(-\Omega t)$.
The markers correspond to eight times
$t=t_0+\slantfrac{1}{8}P_{orb}$,
$t=t_0+\slantfrac{1}{4}P_{orb}$,
$t=t_0+\slantfrac{3}{8}P_{orb}$, $\ldots$,
$t=t_0+P_{orb}$, for which the distributions of
gas--dynamical parameters are presented further in Figs. 4c, 7c,
and 9.
}
\end{figure}

   Under the assumption of uniform rotation, the star should
have the shape of an spheroid that extends to $L_1^{rot}$.
However, the shape calculated using the total potential does not
yield such a solution, since an additional velocity component
normal to the surface arises due to the non-spherical shape of
the star. This implies that the adopted law for the rotation of
the star is not self-consistent.  The problem of determining a
self-consistent rotation law for a star in a binary system
remains unsolved, with the exception of two special cases:
synchronous rotation and the case when the angular-momentum
vector of the donor star is zero in the laboratory frame.
Simultaneously using in the gas-dynamical model the shape of the
stellar surface obtained and the gas-velocity boundary
conditions determined assuming uniform rotation, we obtain a
partially self-consistent solution for this problem. In the
model considered, the gas outflowing from the surface of the
donor star under the adopted boundary conditions can form new
outer layers of the donor star, thereby adjusting the
self-consistent solution. Unfortunately, the presence of
additional forces (the Coriolis force and pressure gradient,
which are not included in the potential) means that we can only
approximate a self-consistent solution; however, this approach
seems optimal for this stage of our study.

\section{CALCULATION RESULTS}

   To analyze the impact of the rotation of the donor star on
the structure of mass flows in the system, we compared models with
synchronous rotation (previous results, presented in [22-25]), with
non-synchronous, aligned rotation, and with non-synchronous, misaligned
rotation of the mass-losing star. We specified the boundary value
of the gas velocity at the surface of the donor star using (6); for both
models with non-synchronous rotation, the rotational velocity of the
star in the laboratory frame was assumed to be twice the angular
rotation velocity of the system. For the calculation with misaligned
rotation, the stellar rotation vector was taken to be inclined to the
$Z$ axis by $15^\circ$ in the laboratory frame (the inclination angle
in the rotating frame was $\vartheta^*=30^\circ$).

\renewcommand{\thefigure}{3}
\begin{figure}[p]
\hbox{\large\hspace{5cm}a)\hspace{6.3cm}b)}
\vspace{1.3cm}
\centerline{\hbox{\psfig{figure=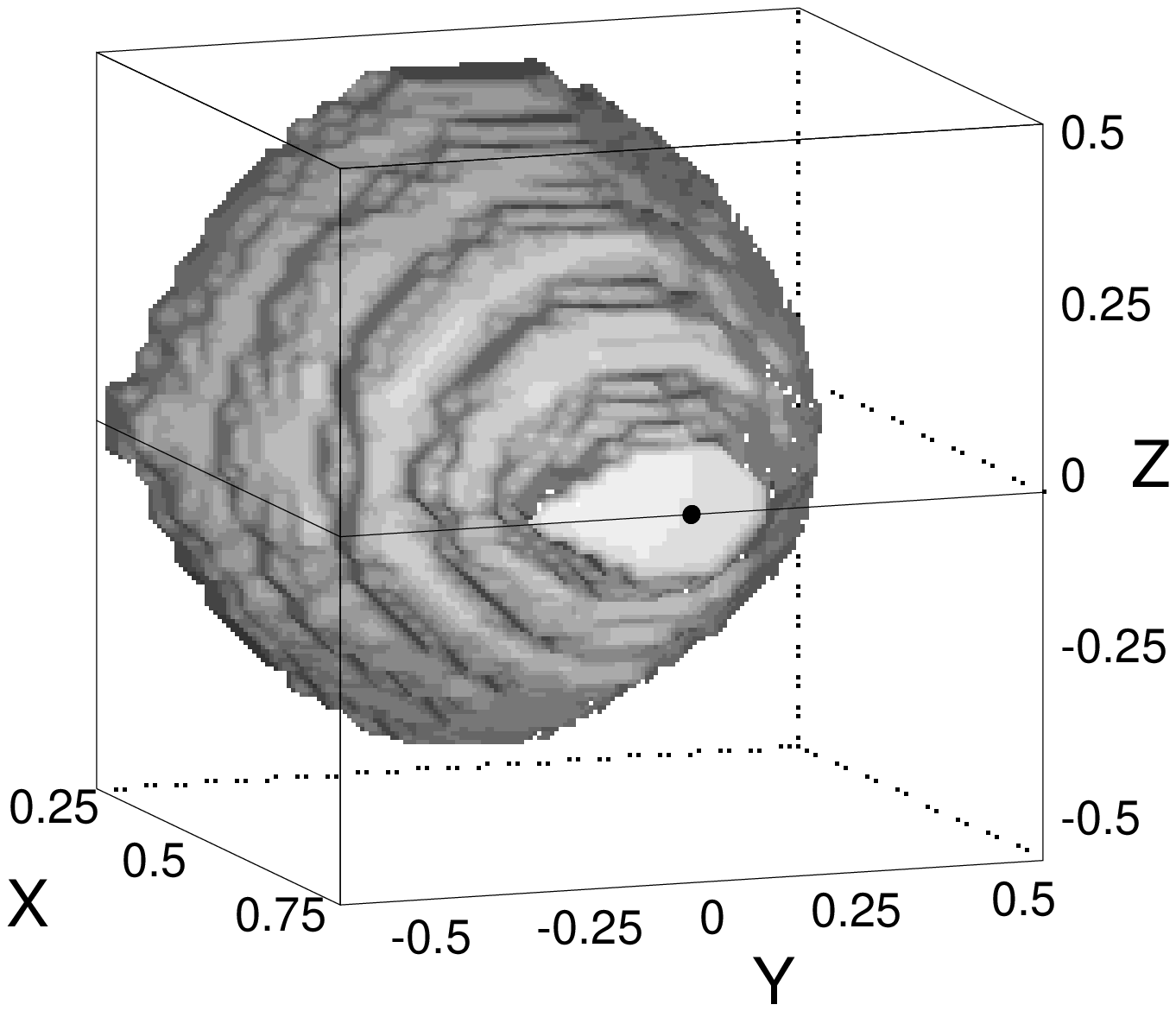,height=2.5in}}
\hspace{1.0cm}
\hbox{\psfig{figure=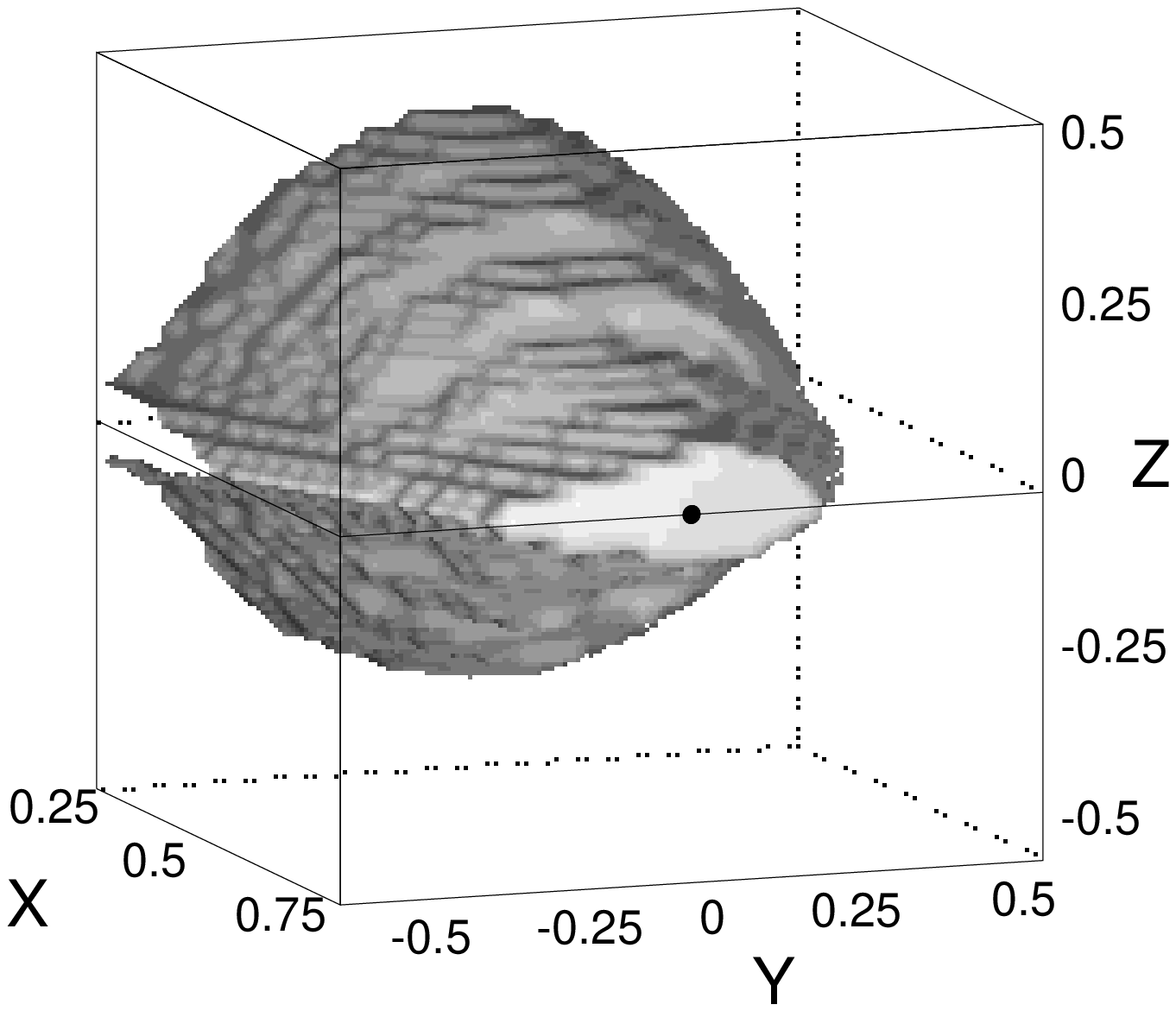,height=2.5in}}}
\vspace{1.0cm}
\hbox{\large\hspace{5cm}c)\hspace{6.3cm}d)}
\vspace{1.3cm}
\centerline{\hbox{\psfig{figure=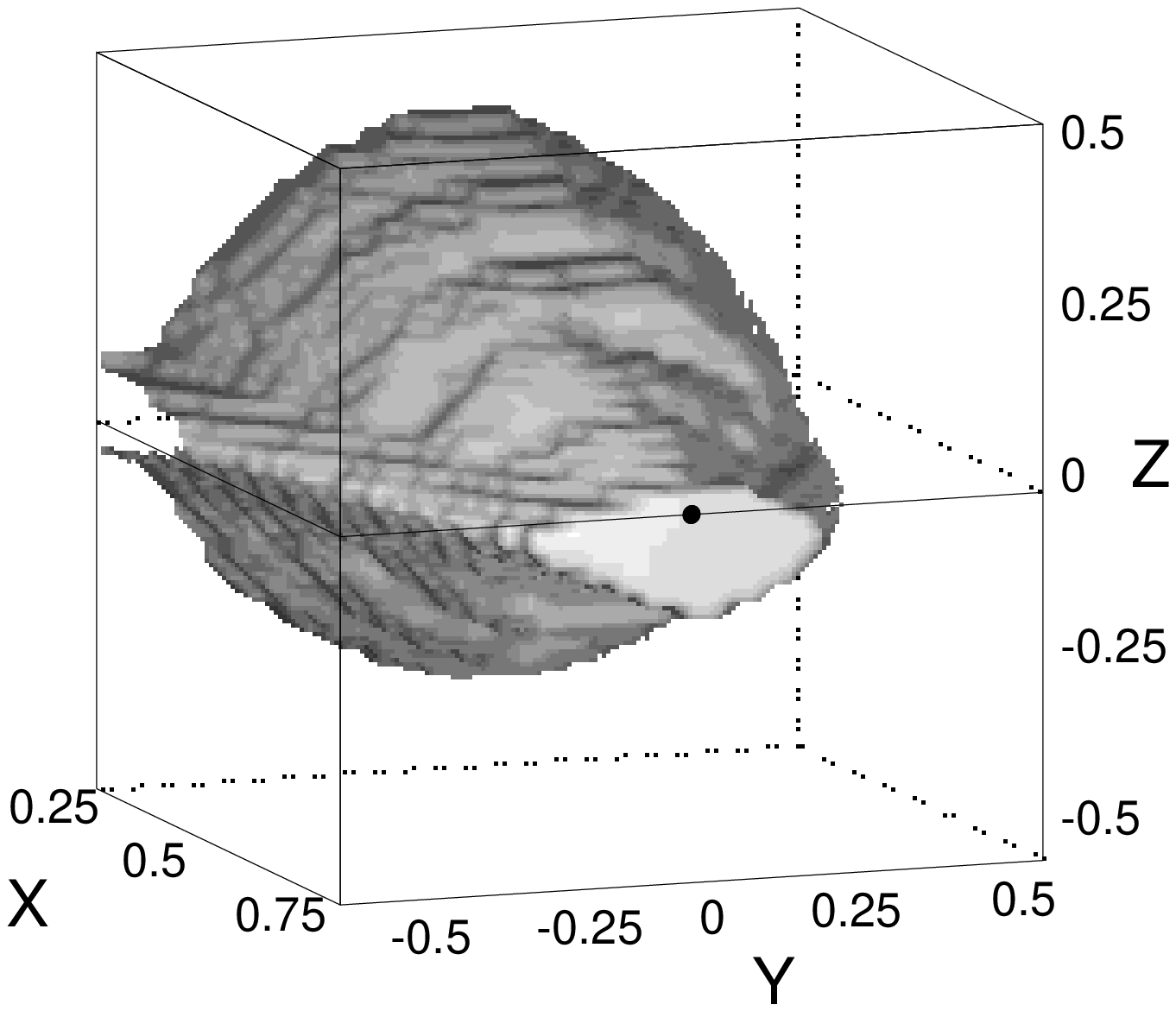,height=2.5in}}
\hspace{1.0cm}
\hbox{\psfig{figure=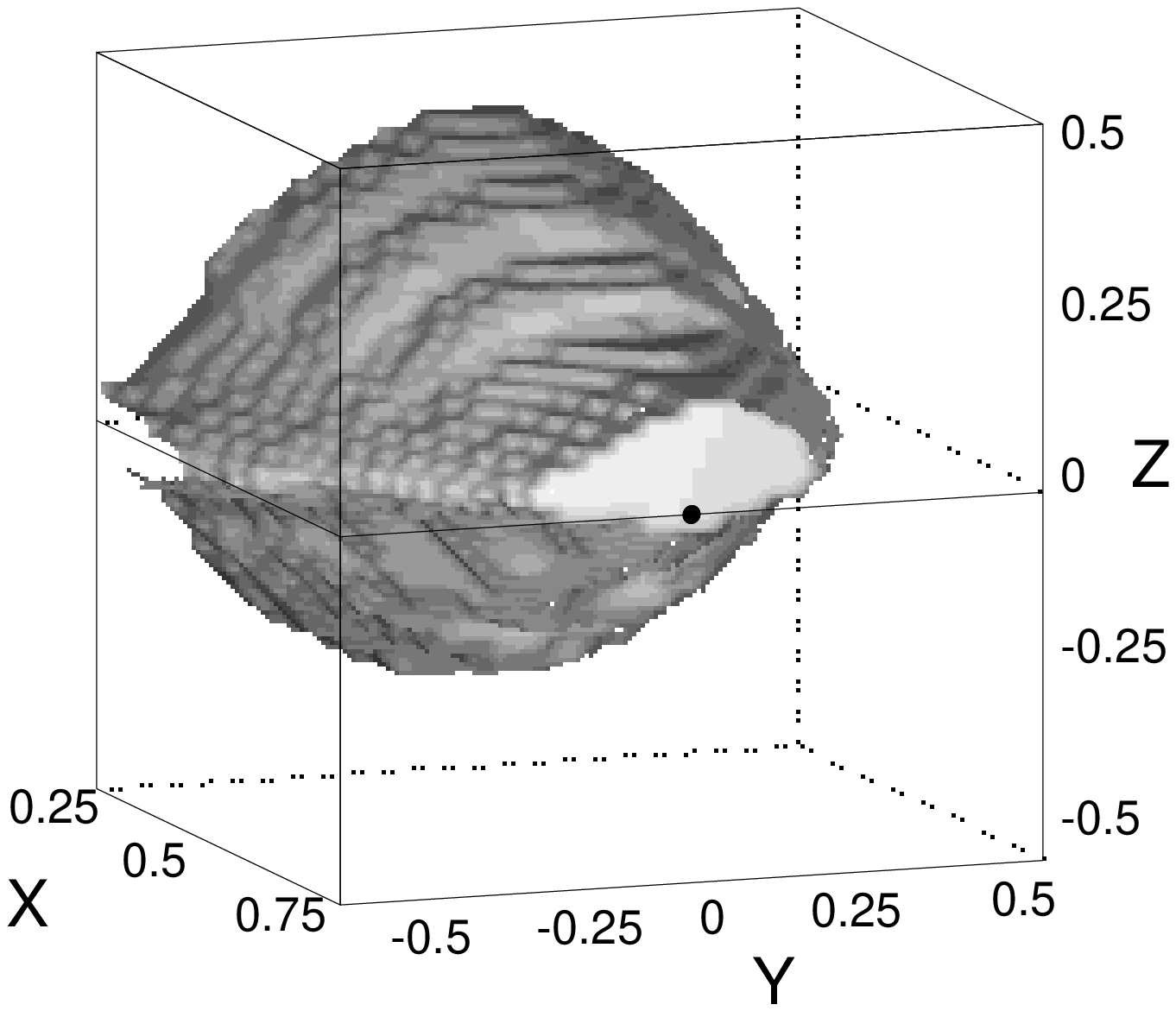,height=2.5in}}}
\caption{\sl
Surface of constant density in the vicinity of the inner
Lagrange point at the level $\rho=0.02\rho_{L_1}$ for the cases
of: (a) synchronous rotation; (b)  non-synchronous, aligned
rotation; (c) and (d) non-synchronous, misaligned rotation for
times $t=t_0+\slantfrac{3}{8}P_{orb}$ and
$t=t_0+\slantfrac{3}{4}P_{orb}$, respectively. The cross
section in the $YZ$ plane was made at a distance of $0.066A$
from $L_1$.  The thick dot marks the projection of the inner
Lagrange point onto the $YZ$ cross section.
}
\end{figure}

\renewcommand{\thefigure}{4}
\begin{figure}[t]
\hbox{\large\hspace{5cm}a)\hspace{7.8cm}b)}
\vspace{0.3cm}
\centerline{\hbox{\psfig{figure=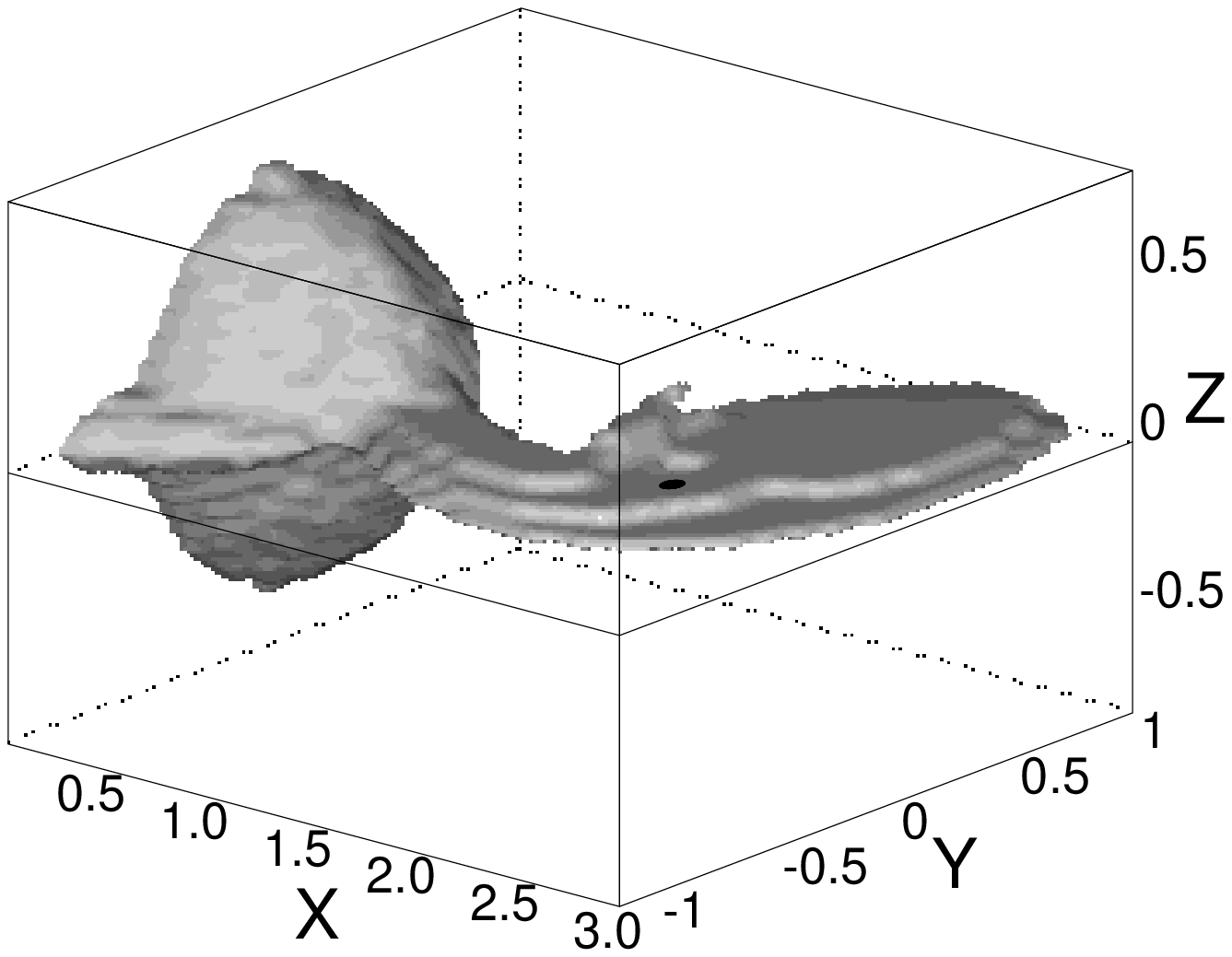,height=2.0in}}
\hspace{1.0cm}
\hbox{\psfig{figure=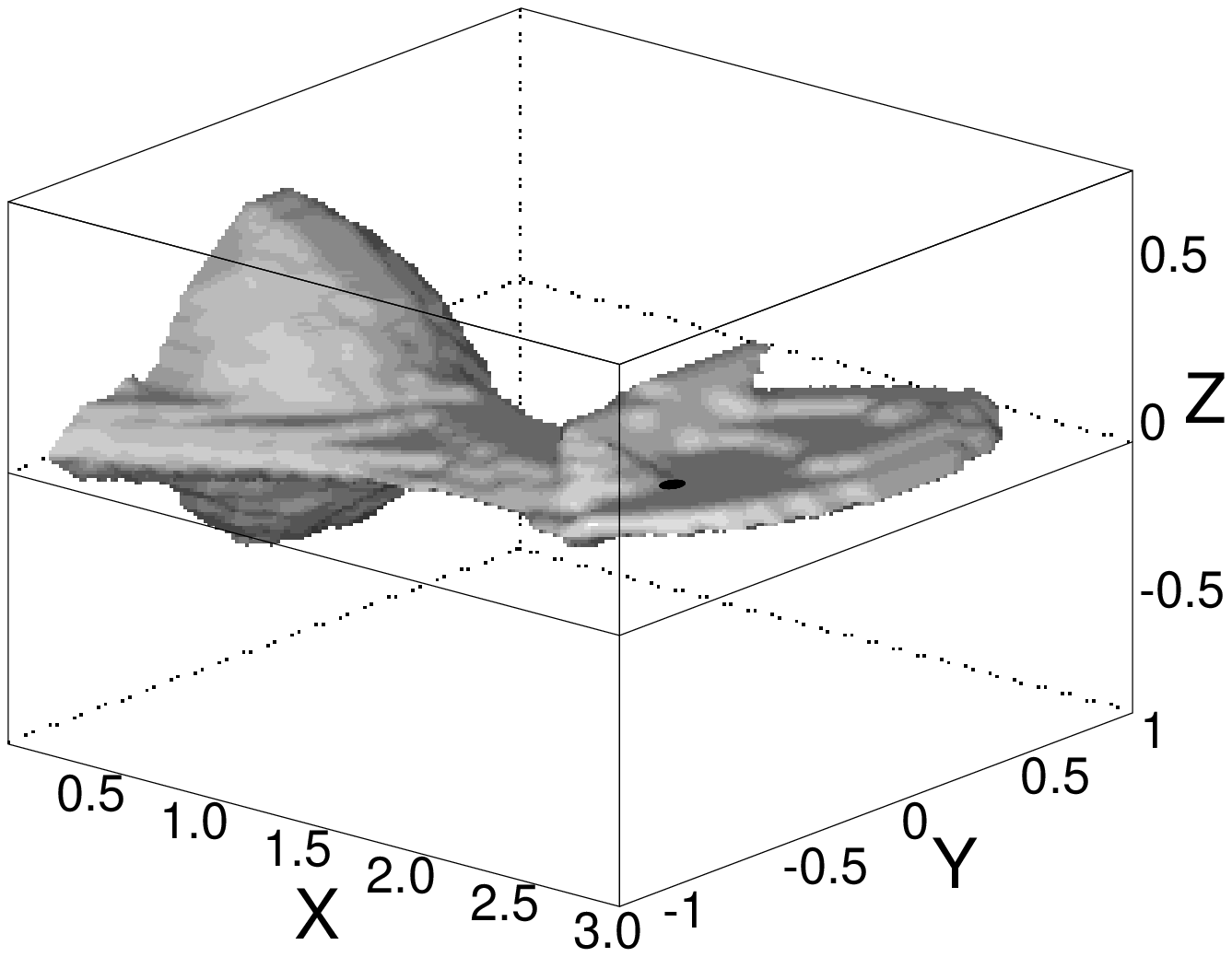,height=2.0in}}}
\caption{\sl
Surface of constant density at the level $\rho=0.004\rho_{L_1}$ for
the case of (1) synchronous rotation and (b) non-synchronous,
aligned rotation. The thick dot marks the position of the
accretor.
}
\end{figure}

\renewcommand{\thefigure}{4c}
\begin{figure}[p]
\centerline{\hbox{\psfig{figure=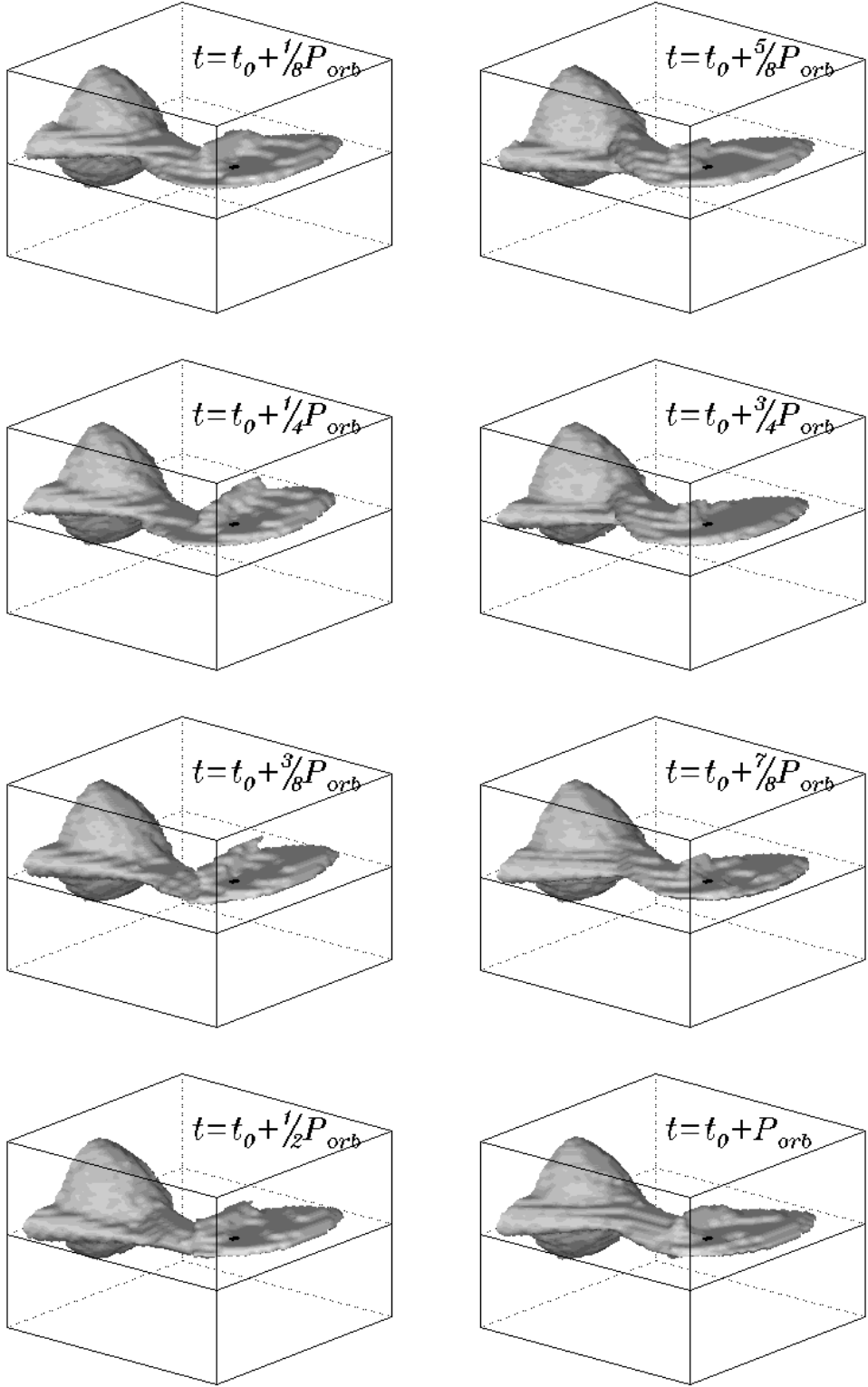,height=8.0in}}}
\caption{\sl
Surface of constant density at
the level $\rho=0.004\rho_{L_1^{rot}}$ for the case of
non-synchronous misaligned rotation for the eight times indicated
in Fig. 2. The size of the area displayed for (c) are the same
as for (a) and (b). The thick dot marks the position of the
accretor.
}
\end{figure}

\renewcommand{\thefigure}{5}
\begin{figure}[t]
\centerline{\hbox{\psfig{figure=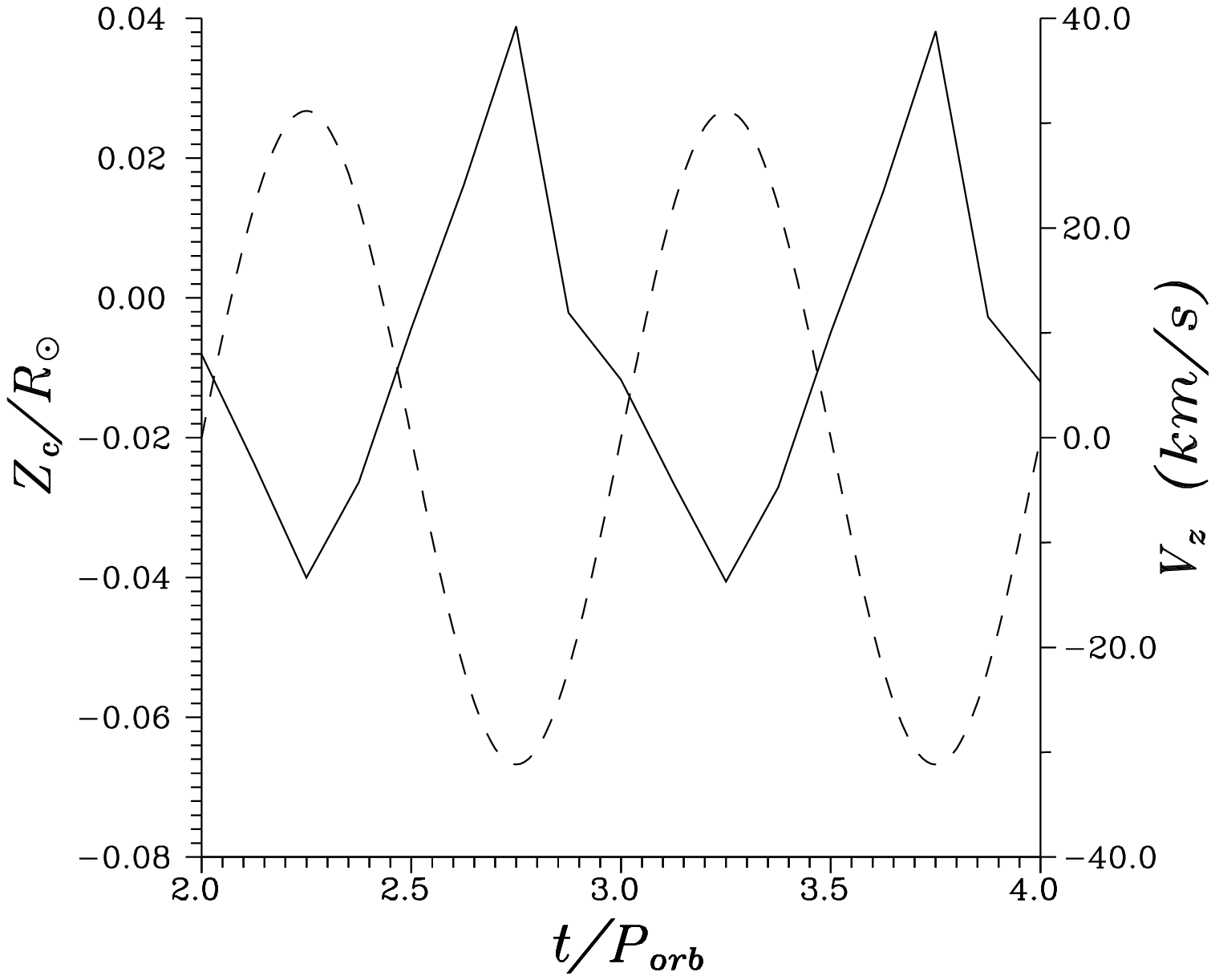,height=3in}}}
\caption{\sl
$Z$ coordinate of the center of mass of the accretion disk
for the case of non-synchronous, misaligned rotation as a function of
time. The dashed line marks the $Z$ component of the velocity at
the inner Lagrange point.
}
\end{figure}

\renewcommand{\thefigure}{6}
\begin{figure}[b]
\caption{(Next page)~\sl
Lines of constant density and velocity vectors in the equatorial
($XY$) plane for the case of (a) synchronous rotation; (b)
non-synchronous, aligned rotation; and (c) non-synchronous,
misaligned rotation for time
$t=t_0+\slantfrac{3}{8}P_{\mbox{®à¡}}$. The thick dot marks the
position of the accretor.  The dashed curves depict the Roche
equipotentials. The vector in the top right corner corresponds
to a velocity of 800 km/s. Stream lines for the flowing matter
are denoted `$a$', `$b$', `$c$'.
}
\end{figure}

\begin{figure}[p]
\hbox{\hspace{12.5cm}\large a)}\vspace{-0.6cm}
\centerline{\hbox{\psfig{figure=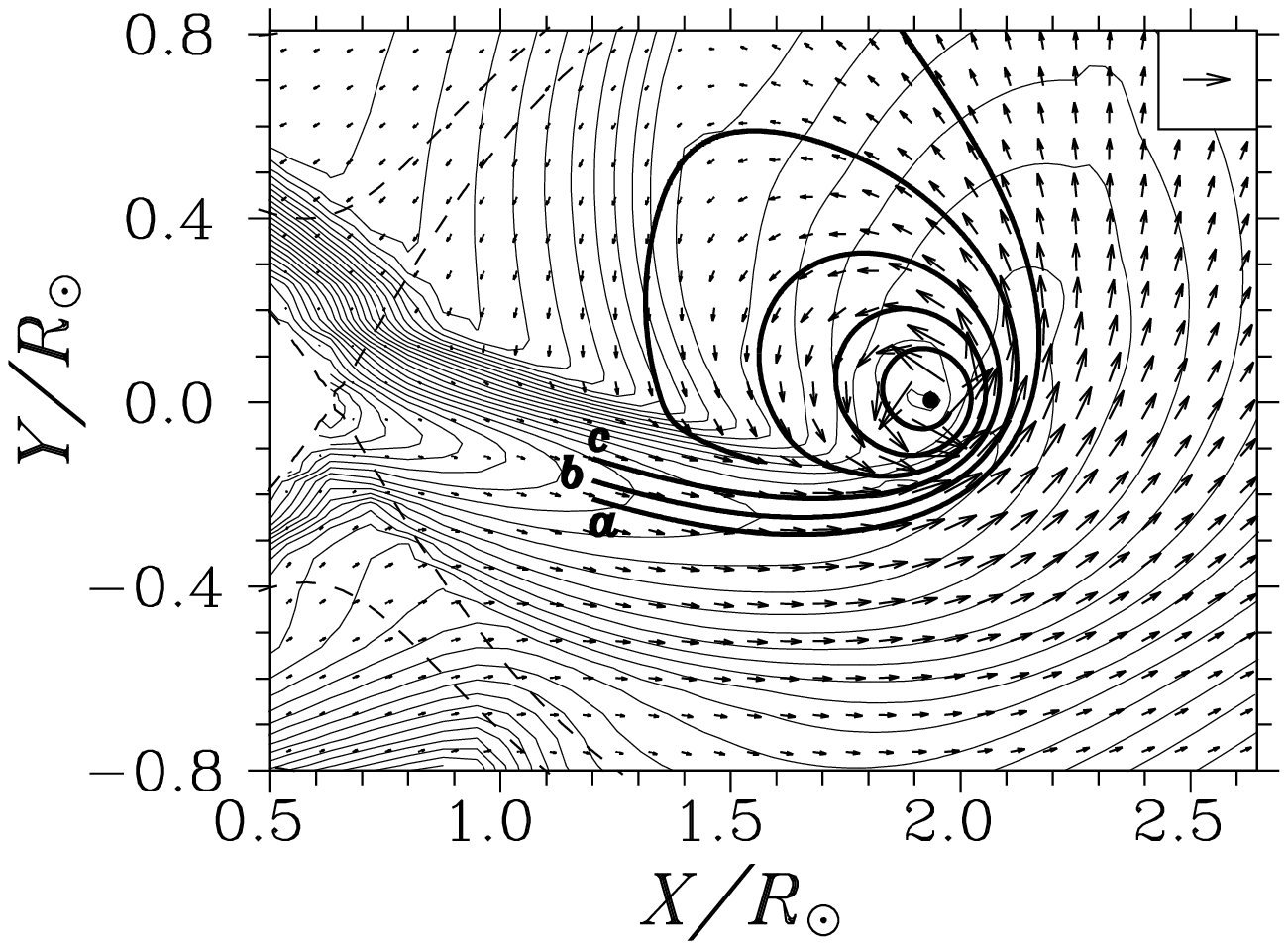,width=4.0in}}}
\hbox{\hspace{12.5cm}\large b)}\vspace{-0.6cm}
\centerline{\hbox{\psfig{figure=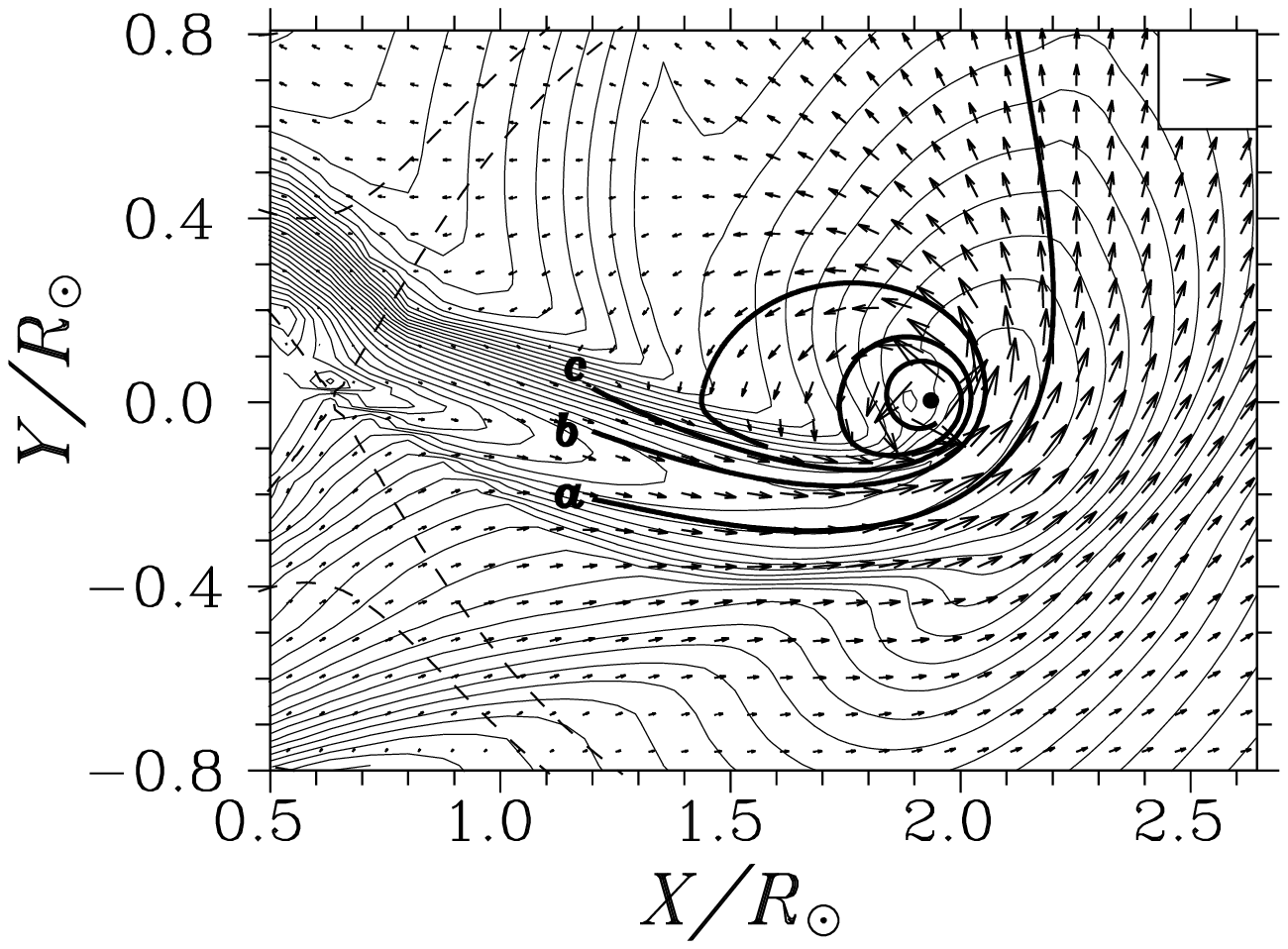,width=4.0in}}}
\hbox{\hspace{12.5cm}\large c)}\vspace{-0.6cm}
\centerline{\hbox{\psfig{figure=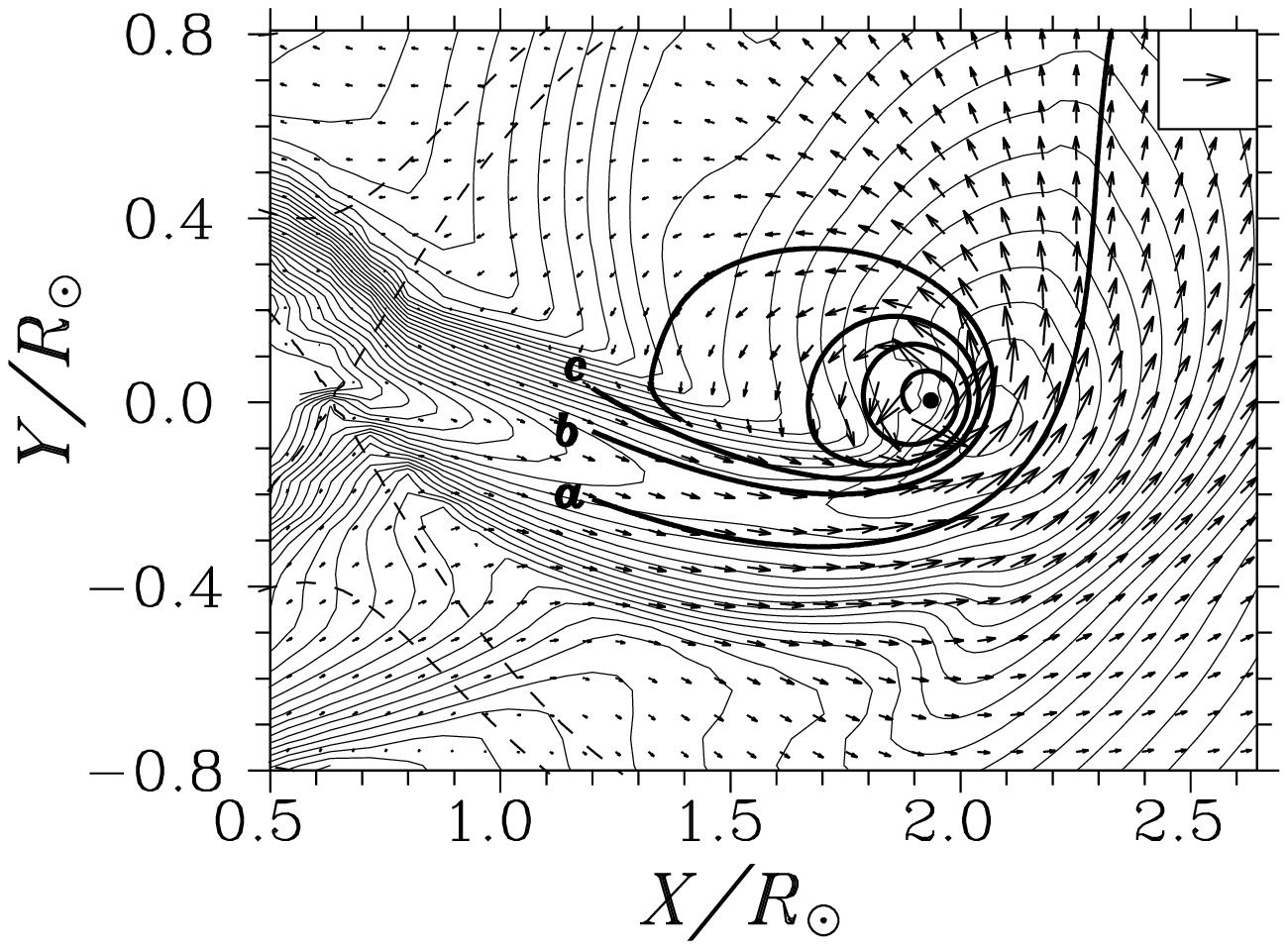,width=4.0in}}}
\end{figure}

\renewcommand{\thefigure}{7}
\begin{figure}[t]
\hbox{\hspace{12.5cm}\large a)}\vspace{-0.6cm}
\centerline{\hbox{\psfig{figure=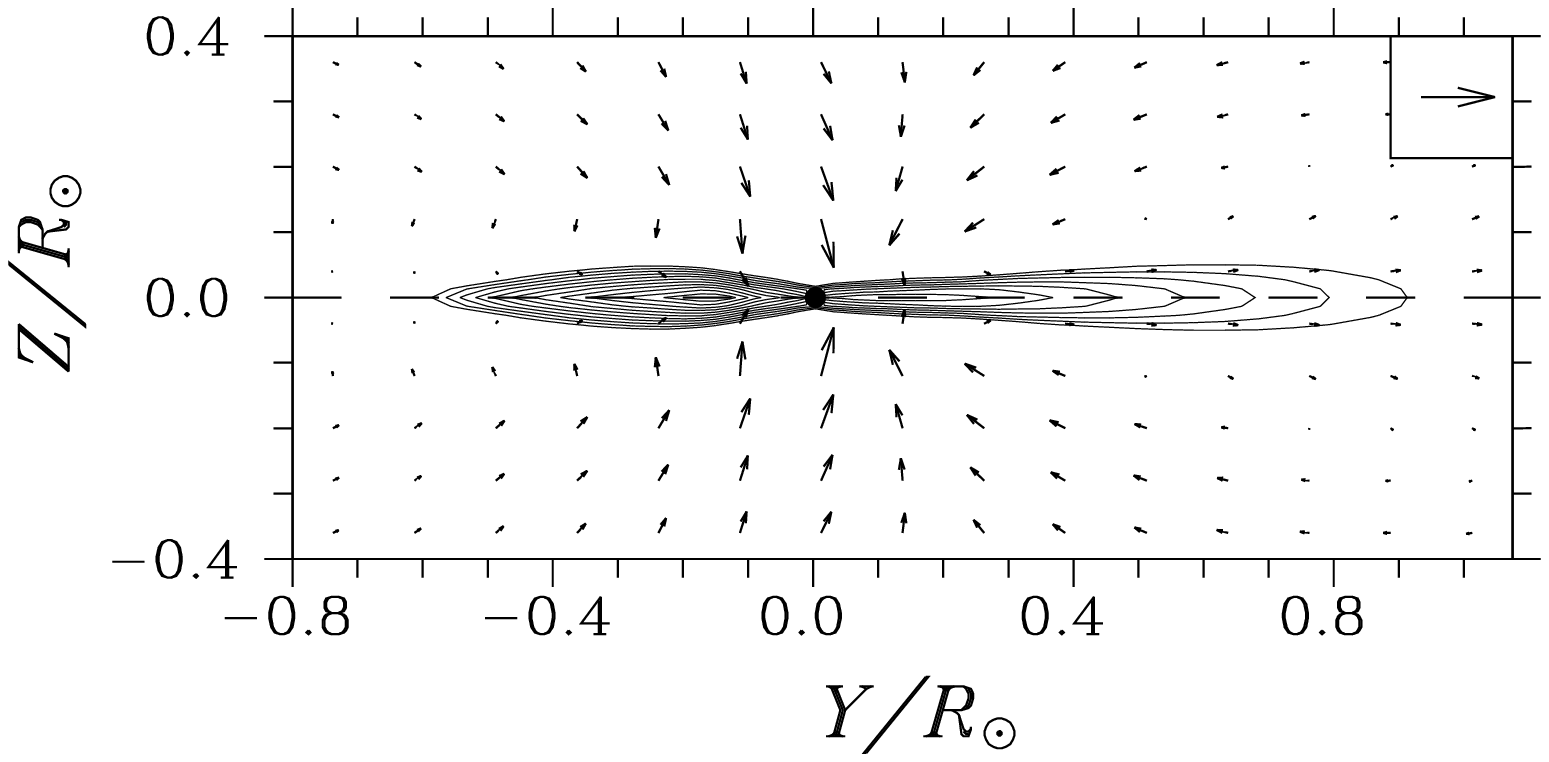,height=2.0in}}}
\hbox{\hspace{12.5cm}\large b)}\vspace{-0.6cm}
\centerline{\hbox{\psfig{figure=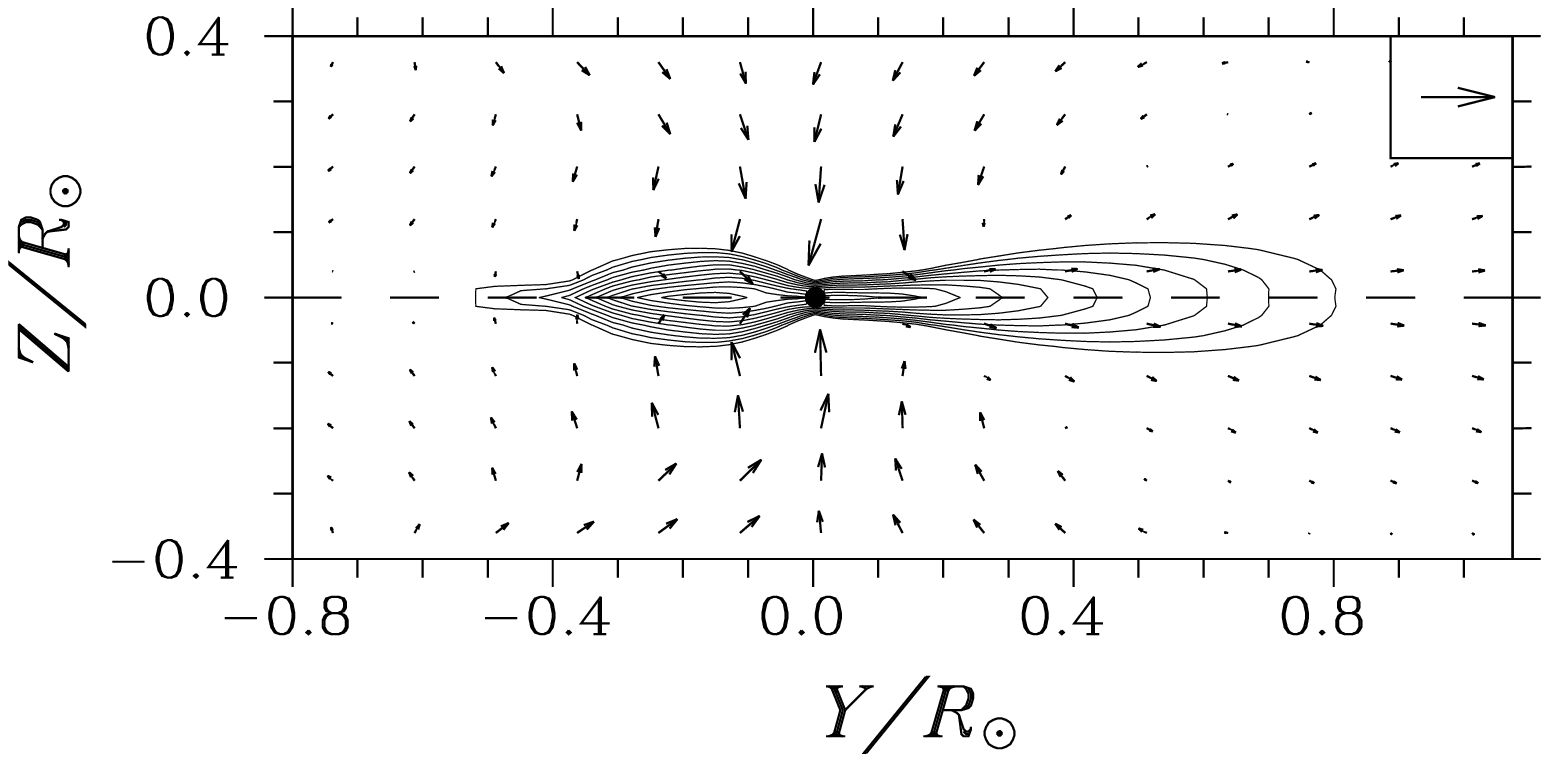,height=2.0in}}}
\caption{\sl
Lines of constant density and velocity vectors in the plane
perpendicular to the line connecting the centers of the stars
(the $YZ$ plane) passing through the accretor for the case of
(a) synchronous rotation, and (b) non-synchronous, aligned
rotation.  The last isoline corresponds to a density of
$\rho=0.005\rho_{L_1}$. The thick dot marks the position of the
accretor. The vector in the top right corner corresponds to a
velocity of 1500 km/s.
}
\end{figure}

\renewcommand{\thefigure}{7c}
\begin{figure}[p]
\centerline{\hbox{\psfig{figure=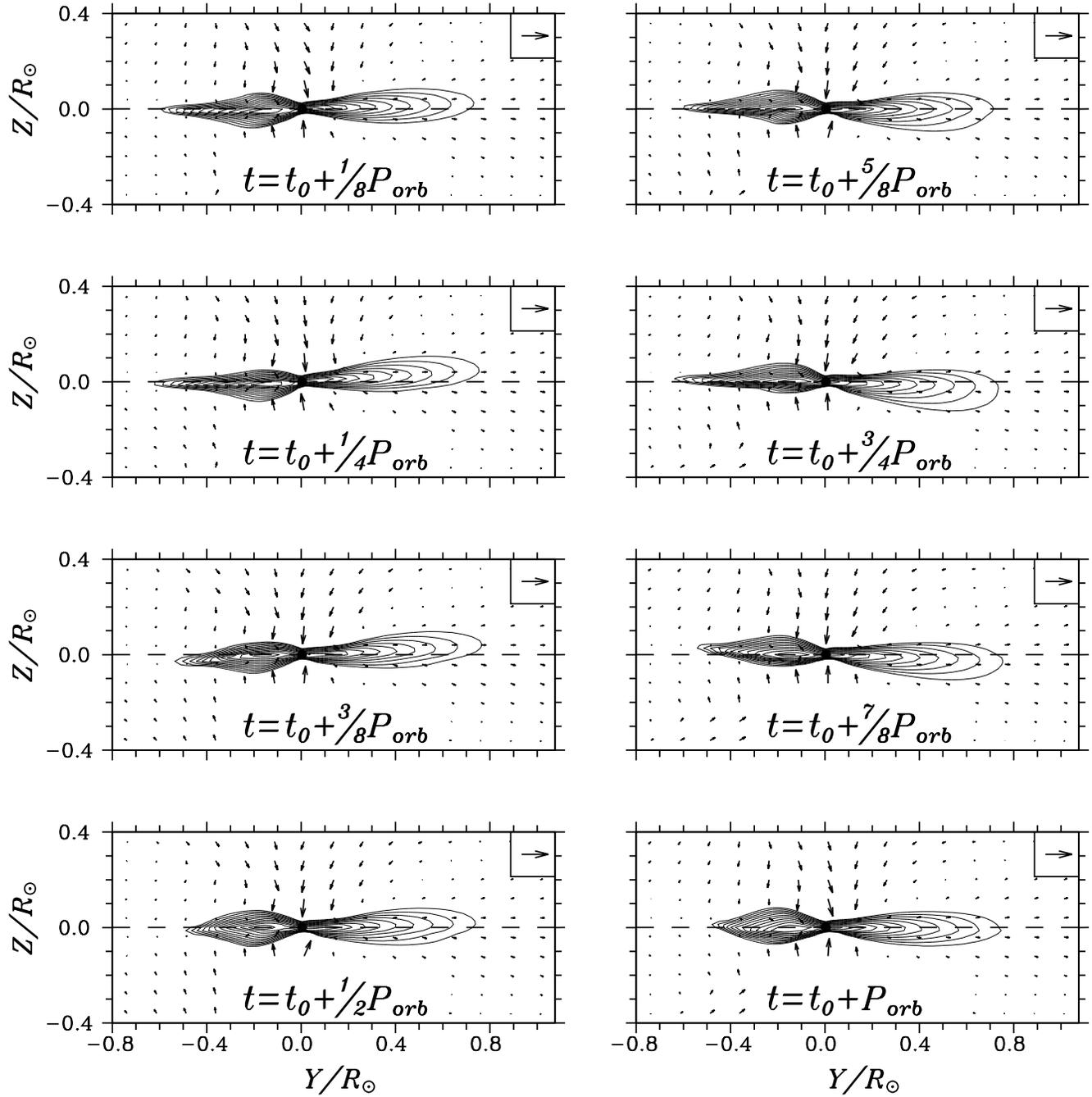,height=7in}}}
\caption{\sl
Lines of constant density and velocity vectors in the plane
perpendicular to the line connecting the centers of the stars
(the $YZ$ plane) passing through the accretor for the case of
non-synchronous, misaligned rotation for the eight times indicated
in Fig. 2. The last isoline corresponds to a density of
$\rho=0.005\rho_{L_1^{rot}}$.
The thick dot marks the position of the
accretor. The vector in the top right corner corresponds to a
velocity of 1500 km/s.
}
\end{figure}

\renewcommand{\thefigure}{8}
\begin{figure}[p]
\hbox{\hspace{11.63cm}\large a)}\vspace{-0.6cm}
\centerline{\hbox{\psfig{figure=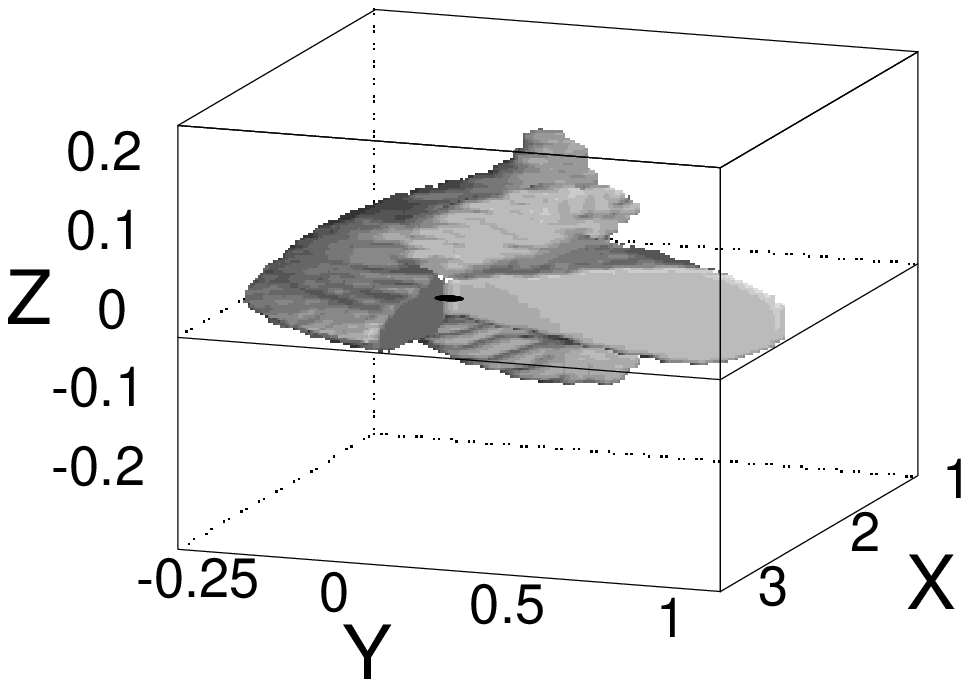,height=2.0in}}}
\vspace{1.0cm}
\hbox{\hspace{11.63cm}\large b)}\vspace{-0.6cm}
\centerline{\hbox{\psfig{figure=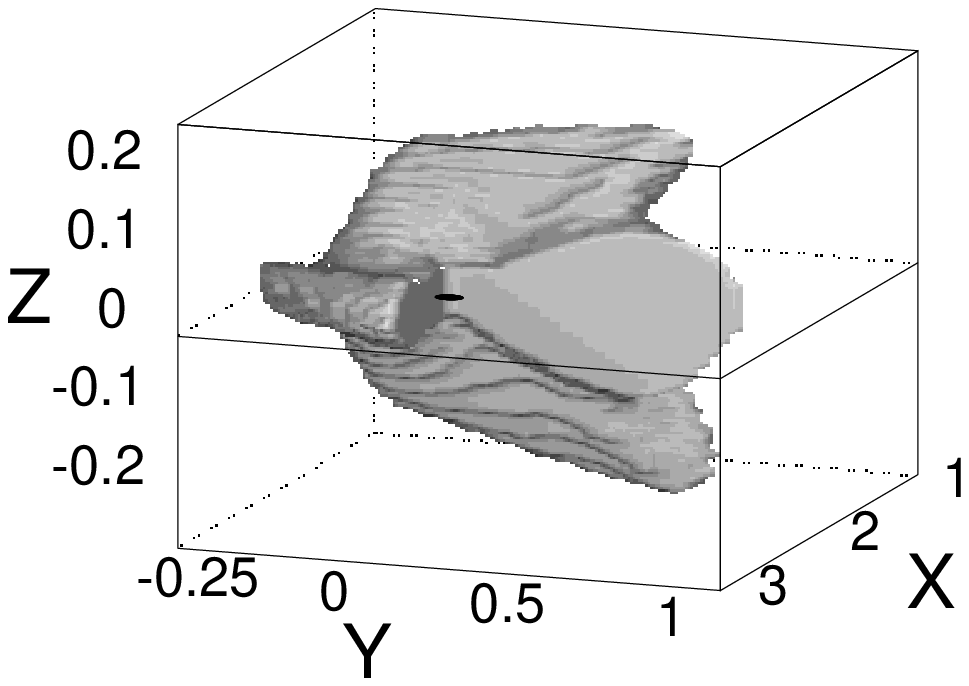,height=2.0in}}}
\vspace{1.0cm}
\hbox{\hspace{11.63cm}\large c)}\vspace{-0.6cm}
\centerline{\hbox{\psfig{figure=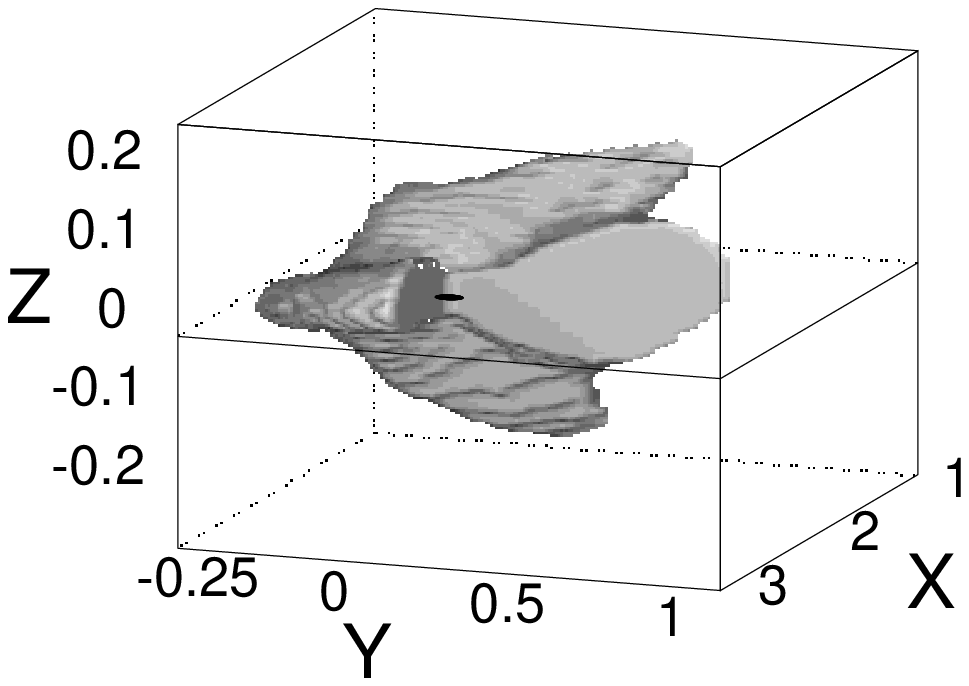,height=2.0in}}}
\caption{\sl
Surface of constant density near the accretor at the level
$\rho=0.004\rho_{L_1}$ for the case of (a) synchronous rotation;
(b) non-synchronous, aligned rotation; and (c) non-synchronous,
misaligned rotation for time
$t=t_0+\slantfrac{3}{8}P_{\mbox{®à¡}}$. The thick dot marks the
position of the accretor. The cross section of the surface made
by the half-planes $y=0~(x>A)$ and $x=A~(y>0)$ also shown.
}
\end{figure}

\renewcommand{\thefigure}{9}
\begin{figure}[p]
\centerline{\hbox{\psfig{figure=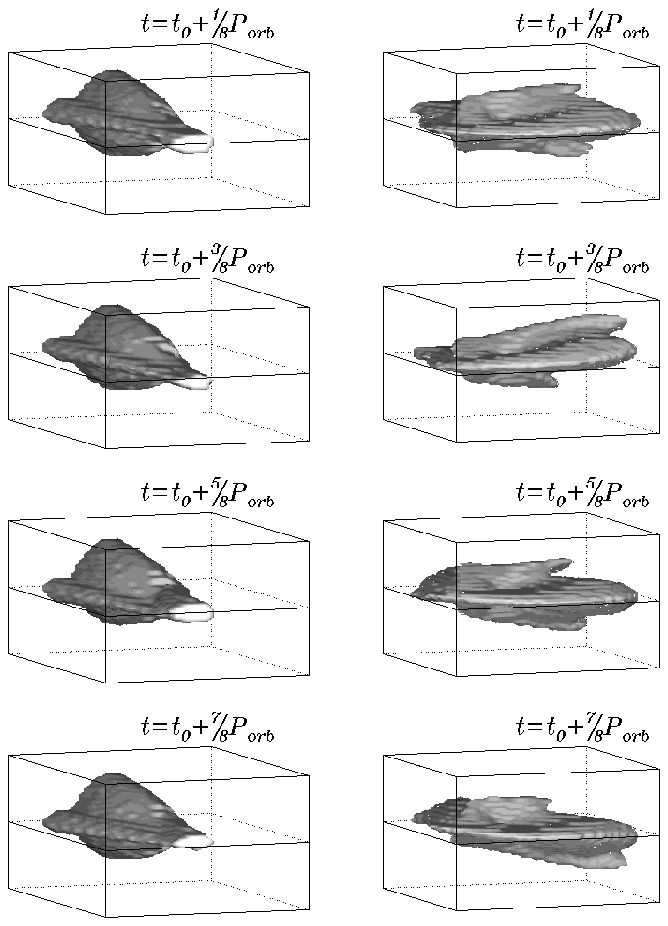,height=7.0in}}}
\caption{\sl
Surfaces of constant density at the level $\rho=0.0025\rho_{L_1^{rot}}$ for
the case of non-synchronous, misaligned rotation at four times in the vicinity
of the inner Lagrange point (left) and the accretion disk (right). The
sizes of the depicted areas are
$[0..5R_\odot]\times[-4R_\odot..4R_\odot]\times[-3R_\odot..3R_\odot]$
(left)
and
$[4R_\odot..11R_\odot]\times[-3R_\odot..4R_\odot]\times[-2R_\odot..2R_\odot]$
(right).
}
\end{figure}

   The different model calculations differ only in the adopted
boundary values for the velocity at the surface of the donor
star. Figure 2 presents the components of the velocity vector at
the inner Lagrange point for all cases. In the adopted reference
frame, the case of synchronous rotation corresponds to the
existence of only the velocity component $V_x$, while $V_y=V_z=0$.
In the case of aligned, non-synchronous rotation, only $V_z=0$,
and in the case of misaligned, non-synchronous rotation all
velocity components are non-zero, with $V_z$ being a periodic
function of time. The differences in the boundary conditions at
the surface of the mass-losing star lead, in particular, to
changes in the parameters for the matter flowing through the
vicinity of the inner Lagrange point, as shown in Fig. 3.
Figures 3a and 3b show surfaces of constant density in the
vicinity of the inner Lagrange point at the level $\rho=0.02\rho_{L_1}$
for the cases of synchronous and aligned non-synchronous rotation.
The cross section of the $YZ$ plane that forms the edge of the
plot in Fig. 3 is close to $L_1$ (at a distance of $0.066A$).
The projection of the inner Lagrange point onto the $YZ$ cross
section is marked by a dot. For the case of misaligned,
non-synchronous rotation, the adopted boundary conditions, and
consequently the solution obtained, are time dependent. Figures
3c and 3d present surfaces of constant density near the inner
Lagrange point at the level $\rho=0.02\rho_{L_1^{rot}}$ for the
two times $t=t_0+\slantfrac{3}{8}P_{orb}$ and
$t=t_0+\slantfrac{3}{4}P_{orb}$, illustrating the change
of the flow parameters, and especially the shift of the flow
relative to the orbital plane. Comparison of Figs. 3a--d shows
that the flow parameters differ substantially for the different
cases, which, in turn, should lead to changes of the flow
patterns obtained for the different types of donor-star
rotation.

   The solutions obtained for synchronous [22--25] and aligned,
non-synchronous rotation are steady-state.  The basic properties of
the obtained flow patterns are shown in Figs. 4a and 4b, which
show surfaces of constant density at the level
$\rho=0.004\rho_{L_1}$ for both model calculations. For the case
of misaligned, non-synchronous rotation, we considered the
solution to be steady-state when the main features of the flow
structure repeated with the period of the boundary conditions.
Figure 4c presents surfaces of constant density at the level
$\rho=0.004\rho_{L_1^{rot}}$ for eight times (Fig. 2) covering
the total variation period for the boundary conditions.
Analysis of the results presented in Fig. 4c shows that the
behavior of the matter in and around the disk reflects changes
in the injected matter (as well as changes in the boundary
conditions at the surface of the donor star). This implies that
a driven-disk model has been realized in our calculations.  An
additional illustration of this fact is provided by Fig. 5,
which shows the temporal dependencies of variations in the
$Z$-coordinates of the center of mass of the accretion disk and
of the velocity at the inner Lagrange point.  Analysis of the
curves presented in Fig. 5 confirms the "driven" character of
the solution, and the strict dependence of the flow pattern near
the accretor on the behavior of the flow in the vicinity of
$L_1$.  Note that the inclination angle of the normal to the
disk surface to the $Z$ axis is $\pm 10^\circ$ in the $XZ$ plane
and $\pm 20^\circ$ in the $YZ$ plane.

   Let us consider the details of the flow pattern obtained. For
the cases of synchronous and aligned, non-synchronous rotation, the
main features of the steady-state flow pattern obtained are
clearly visible in Figs. 4a,b and 6a,b. Our results indicate
that the rarefied gas of the circumbinary envelope (stream
lines `$a$' and `$b$') has an appreciable impact on the
structure of gas flows in the system. The gas of the
circumbinary envelope interacts with the matter flowing from
the vicinity of $L_1$ and deflects it, leading to a shock-free
(tangential) interaction of the flow with the outer edge of the
forming accretion disk (stream line `$c$'), and, consequently,
to the absence of a "hot spot" on the disk. At the same time,
the interaction of the circumbinary-envelope gas (stream line
`$b$') with the flow leads to the formation of an extended shock
of variable intensity along the edge of the flow. The impact of
aligned, non-synchronous rotation of the donor star is manifested
only in a number of quantitative changes, while the overall flow
pattern remains qualitatively similar.

   As noted above, in the case of non-synchronous rotation, the
existence of a $V_y$ component for the velocity brings about
changes in the parameters of the inflowing matter (Fig. 3).
Because of the shift of the flow near $L_1$, the flow approaches
the accretor as it advances, and the disk that forms is
substantially smaller than in the case of synchronous rotation
(Figs. 6a,b). In our calculation for non-synchronous rotation, the
thicknesses of the flow, the accretion disk, and the structures
surrounding the disk exceed the disk thickness in the
synchronous case (Figs. 7a,b; 8a,b). Note that, in addition to
forming a shock, the interaction of the rarefied gas of the
circumbinary envelope (stream line `$b$' in Fig. 6) with the
flow also leads to the formation of gaseous "clouds" that
precede the front edge of the flow outside the disk-formation
area (Figs. 4, 8). The appearance of these formations is
associated with the interaction (collision) of rarefied envelope
gas outside the equatorial plane with the flow.  The larger
thickness of the flow in the case of aligned, non-synchronous
rotation brings about the formation of more pronounced "clouds".

  In the calculations for misaligned, non-synchronous rotation,
all the basic features of the flow inherent to our previous
calculations are maintained.  Characteristic properties, such as
the formation of an circumbinary envelope, the absence of a
"hot spot" at the edge of the accretion disk, and the formation
of a shock wave along the edge of the flow, are clearly visible
in Figs. 4c and 6c. As noted above, a "driven disk" is realized
in this case, and there are periodic changes in the solution in
the steady-state regime (Figs. 4c, 5, 7c). Unlike the case of
aligned, non-synchronous rotation, the matter in the disk and the
surrounding structures fluctuate about the equatorial plane,
leading to the formation of complicated flow structures
("tracks"). The presence of these "tracks" of matter means that
the disk that forms possesses intermediate linear size and
thickness, between those for the models with synchronous and
aligned, non-synchronous rotation (Figs. 6, 7, 8). Due to the
"driven" character of the solution, the formation of "clouds"
(gaseous formations above and below the equatorial plane) is
also periodic (Fig. 9). Since it is not excluded that these
formations can contribute to the total emission of the system,
together with the emission of the disk and surrounding material,
the periodicity of their formation should be considered when
interpreting observations.

\section*{CONCLUSIONS}

   We have presented the results of three-dimensional numerical
modeling of the mass transfer in semi-detached binary systems
with rotation of the mass-losing star. We considered the cases
of both aligned and misaligned, non-synchronous rotation of the
donor star, and compared the flow patterns obtained with the
synchronous case.

   Our analysis indicates that the main features of the flow are
qualitatively similar for all the calculations. Study of the
flow structure indicates that the rarefied gas of the
circumbinary envelope has an appreciable impact on the
structure of gas flows in the system. The
circumbinary-envelope gas interacts with the matter flowing
from the vicinity of $L_1$ and deflects it, leading to a
shock-free (tangential) interaction of the flow with the outer
edge of the forming accretion disk, and, consequently, to the
absence of a "hot spot" on the disk. At the same time, the
interaction of this gas with the flow forms an extended shock of
variable intensity along the edge of the flow.

   Unlike the cases of synchronous and aligned, non-synchronous
rotation of the donor star, in the case of misaligned,
non-synchronous rotation, it is impossible to achieve a genuine
steady-state regime due to the periodic time dependence of the
boundary conditions. Therefore, in the misaligned, non-synchronous
case, the solution was considered to be steady-state when the
basic features of the flow structure repeated with the period of
the boundary conditions. Our analysis indicates that the
behavior of the disk and surrounding material reflects changes
in the injected matter (or changes in the boundary
conditions at the surface of the donor star). This implies that
a "driven disk" model has been realized in our calculations. The
inclination angle of the normal vector to the disk surface to
the $Z$ axis is $\pm 10^\circ$ in $XZ$ plane and $\pm 20^\circ$
in the $YZ$ plane.  The calculated periodic changes of the shape
of the accretion disk and of the surrounding gaseous envelope
should be observable via the associated changes in the emission
of this region.

\section*{ACKNOWLEDGMENTS}
   This work was supported by the Russian Foundation for Basic
Research (project code 96-02-16140) and the INTAS Foundation
(grant 93-93-EXT).

\section*{REFERENCES}

\begin{enumerate}

\item Tassoul, J.-L. 1978 {\it Theory of rotating stars},
Princeton Univ. Press, Princeton

\item Wilson, R.E., Van Hamme, W., \& Petera, L.E. 1985,
ApJ, 289, 748

\item Van Hamme, W., \& Wilson, R.E. 1992, AJ, 92, 1168

\item Habets, G.M.H.J., \& Zwaan, C. 1989, A\&A, 211, 56

\item Griffin, R., \& Griffin, R. J. 1986, A\&A, 7, 45

\item Fekel, F.C., Moffet, T.J., \& Henry, G.W. 1986, ApJS, 60,
551

\item Shcherbakov, A.G., Tuyominen, I., Jetsu, L., Katsova M.M.,
\& Poutanen M. 1990, A\&A, 235, 205

\item Ayres, T.R. 1991, ApJ, 375, 704

\item Katsova, M.M. 1990, Lect. Notes Phys., 397, 220

\item Shakura, N.I. 1972, Astron. Zh., 49, 921
(Sov. Astron., 16, 756)

\item Roberts, W.J. 1974, ApJ, 187, 575

\item Petterson, J.A. 1975, ApJ, 201, L61

\item Katz, J.I. 1979, ApJ, 236, L127

\item Shklovskii, I.S. 1979, Pis'ma Astron. Zh., 5, 644
(Sov. Astron. Lett., 5, 344)

\item Van den Heuvel, E.P.J., Ostriker, J.P., \& Petterson,
J.A. 1980, A\&A, 81, 27

\item Whitmire, D.P., \& Matese, J.J. 1980, MNRAS, 193, 707

\item Cherepashchuk, A.M. 1981, Pis'ma Astron. Zh., 7, 201
(Sov. Astron. Lett., 7, 111)

\item Kruszewski, A. 1964, Acta Astron., 14, 231

\item Belvedere, G., Lanzafame, G., \& Molteni, D.,
1993, A\&A, 280, 525

\item Lanzafame, G., Belvedere, G., \& Molteni, D. 1994, MNRAS,
267, 312

\item Avni, Y., \& Schiller, N. 1982, ApJ, 257, 703

\item Bisikalo, D.V., Boyarchuk, A.A., Kuznetsov, O.A., \&
Chechetkin, V.M. 1997, Astron. Zh., 74, 880
(Astron. Reports, 41, 786; preprint astro-ph/9802004)

\item Bisikalo, D.V., Boyarchuk, A.A., Kuznetsov, O.A., \&
Chechetkin, V.M. 1997, Astron. Zh., 74, 889
(Astron. Reports, 41, 794; preprint astro-ph/9802039)

\item Bisikalo, D.V., Boyarchuk, A.A., Chechetkin, V.M.,
Molteni D., \& Kuznetsov O.A. 1998, MNRAS, 300, 39
(preprint astro-ph/9805261)

\item Bisikalo, D.V., Boyarchuk, A.A., Kuznetsov, O.A., \&
Chechetkin, V.M. 1998, Astron. Zh., 75, 706
(Astron. Reports, 42, 706; preprint astro-ph/9806013)

\item Armitage, P.J., \& Livio, M. 1996, ApJ, 470, 1024

\item Kopal, Z. 1978, {\it Dynamics of close binary systems},
Reidel, Dordrecht

\item Pringle, J.E., \& Wade, R.A. (Eds.) 1985, {\it Interacting
binary stars} Cambridge Univ. Press,  Cambridge

\item Plavec, M. 1958, M\'em. Soc. Roy. Sci. Li\`ege, 20, 411

\item Kruszewski, A. 1963, Acta Astron., 13, 106

\item Limber, D.N. 1963, ApJ, 138, 1112

\item Kruszewski, A. 1966, Adv. Astron. Astrophys., 4, 233

\item Layton, J.T., Blondin, J.M., Owen, M.P., \& Stevens, I.R.
1998, New Astron., 3, 111

\item Lubow, S.H., \& Shu, F.H. 1975, ApJ, 198, 383

\item Sawada, K., Matsuda, T., \& Hachisu, I. 1986, MNRAS, 219, 75

\item Bisikalo, D.V., Boyarchuk, A.A., Kuznetsov, O.A.,
Chechetkin V.M. 1995, Astron. Zh., 72, 367
(Astron. Reports, 39, 325)

\item Roe, P.L. 1986, Ann. Rev. Fluid. Mech., 18, 337

\item Osher, S., \& Chakravarthy, S. 1984, SIAM J. Numer.
Anal., 21, 955

\end{enumerate}


\end{document}